\documentclass[manuscript,screen]{acmart}

\usepackage{longtable}
\usepackage{booktabs}
\usepackage{tabularx}
\usepackage{array}
\usepackage{multirow}
\usepackage{makecell}
\usepackage{placeins}
\usepackage[table]{xcolor}

\AtBeginDocument{%
  }

\acmDOI{} 

\begin{document}

\title[Trustworthy AI in Fintech]{When AI Meets Wall Street: A Survey on Trustworthy AI in Fintech}

\author{Qingwen Zeng}
\authornote{These authors contributed equally to this research.}
\orcid{0009-0002-7926-3606}
\email{qzen5227@uni.sydney.edu.au}
\affiliation{%
  \institution{School of Electrical and Computer Engineering, The University of Sydney}
  \city{Camperdown, NSW 2006}
  \country{Australia}
}

\author{Zhenghao Zhao}
\authornotemark[1]
\orcid{0009-0000-2287-6423}
\email{zzha0595@uni.sydney.edu.au}
\affiliation{%
  \institution{School of Electrical and Computer Engineering, The University of Sydney}
  \city{Camperdown, NSW 2006}
  \country{Australia}
}

\author{Yitian Yang}
\authornotemark[1]
\orcid{}
\email{yyan5712@uni.sydney.edu.au}
\affiliation{%
  \institution{School of Electrical and Computer Engineering, The University of Sydney}
  \city{Camperdown, NSW 2006}
  \country{Australia}
}

\author{Yiqi Zhu}
\orcid{}
\email{yzhu0325@uni.sydney.edu.au}
\affiliation{%
  \institution{School of Computer Science, The University of Sydney}
  \city{Camperdown, NSW 2006}
  \country{Australia}
}

\author{Fangchen Liu}
\orcid{}
\email{fliu6119@uni.sydney.edu.au}
\affiliation{%
  \institution{School of Electrical and Computer Engineering, The University of Sydney}
  \city{Camperdown, NSW 2006}
  \country{Australia}
}

\author{Zhaoge Bi}
\orcid{}
\email{zhaoge.bi@sydney.edu.au}
\affiliation{%
  \institution{School of Electrical and Computer Engineering, The University of Sydney}
  \city{Camperdown, NSW 2006}
  \country{Australia}
}

\author{Moe Thandar Kyaw Wynn}
\email{m.wynn@qut.edu.au}
\orcid{0000-0002-7205-8821}
\affiliation{%
  \institution{Queensland University of Technology}
  \city{Brisbane City QLD 4000}
  \country{Australia}
}

\author{Kim-Kwang Raymond Choo}
\email{raymond.choo@fulbrightmail.org}
\orcid{0000-0001-9208-5336}
\affiliation{
  \institution{University of Texas at San Antonio}
  \city{San Antonio}
  \state{TX}
  \country{USA}
}
\author{Huaming Chen}
\authornote{Corresponding author.}
\orcid{0000-0001-5678-472X}
\email{huaming.chen@sydney.edu.au}
\affiliation{%
  \institution{School of Electrical and Computer Engineering, The University of Sydney}
  \city{Camperdown, NSW 2006}
  \country{Australia}
}

\renewcommand{\shortauthors}{Zeng, et al.}

\begin{abstract}
Artificial intelligence is now embedded as a primary decision engine in continuously operated financial AI pipelines spanning training and updating, deployment and inference, and operation with monitoring and feedback. The automation and scale that make these pipelines effective also create novel attack surfaces, where small algorithmic perturbations can amplify into persistent, system-level financial harm. Existing surveys, however, either treat AI as a defensive tool or analyse adversarial machine learning in a domain-agnostic manner, abstracting away finance-specific constraints such as accounting plausibility, non-IID federated data, continuous retraining, and automation-amplified downstream effects. We address this gap with a unified, lifecycle-centric and mechanism-driven framework. We partition financial AI into three lifecycle stages: training and updating, deployment and inference, and operation, monitoring, and feedback. We further propose the Financial AI Security \& Robustness Taxonomy, organising seventeen attack subtypes across data and model poisoning, adversarial attacks on decision boundaries, prompt injection in LLM-mediated workflows, and deepfake-driven subversion of KYC verification layers. For each subtype, we analyse algorithmic strategy, feasibility constraints, stealth and persistence, and downstream financial consequences. Finally, we identify open challenges and outline a research agenda toward lifecycle-aware stress testing and finance-relevant robustness benchmarks.
\end{abstract}

\keywords{financial AI, fintech, trustworthy machine learning, model robustness, LLM agents, fraud detection, risk management}

\maketitle

\section{Introduction}
\label{sec:introduction}

Artificial intelligence (AI) has become a foundational component of modern financial systems, supporting automated decision-making across a diverse range of tasks, i.e., credit scoring, fraud and anti-money laundering/counter terrorism financing (AML/CTF), algorithmic trading, and risk management~\cite{bahoo2024artificial,kou2025fintech}. In contrast to earlier deployments where machine learning models served as auxiliary analytic tools, contemporary financial AI systems increasingly operate as primary decision engines embedded within end-to-end pipelines that integrate data collection, feature engineering, model training and updating, inference-time decision execution, and continuous monitoring~\cite{cao2024applied, brahmandam2025mlops}. These pipelines are typically deployed at scale, operate with minimal human intervention, and are tightly coupled to downstream automated actions~\cite{aldasoro2025predicting}. As a result, financial AI systems should not be viewed as isolated models, but rather as continuously operated, automation-amplified socio-technical pipelines whose behaviour emerges from the interaction between learning algorithms, deployment mechanisms, and operational feedback loops~\cite{pattnaik2024applications}. 

A defining characteristic of such pipelines is that small algorithmic effects can be amplified into persistent system-level consequences. Financial AI systems often operate within extended decision chains, where model outputs trigger subsequent automated processes, such as transaction blocking, credit approval, portfolio rebalancing, or risk threshold adjustment, with limited opportunities for human oversight~\cite{weber2024applications, calvano2020artificial}. In this setting, minor deviations introduced during training, inference, or updating can propagate through automation and persist via feedback mechanisms, resulting in disproportionate and long-lived impacts~\cite{pattnaik2024applications,bahoo2024artificial,perdomo2020performative,d2022underspecification}. From a technical perspective, this amplification arises not from isolated software failures but from the interaction between learning dynamics, optimisation procedures, and deployment architectures that continuously reuse model outputs as future inputs~\cite{shumailov2023curse}. Understanding security and robustness in financial AI, therefore, requires a lifecycle-centric analysis that accounts for how algorithmic perturbations emerge, propagate, and persist across system stages. 

At the same time, the same properties that make financial AI pipelines operationally effective, including automation, scale, and continuous interaction, also create novel attack surfaces~\cite{board2024financial, chen2024security,lai2024security}. Unlike conventional software systems, adversaries in financial environments often operate through legitimate interaction channels: shaping effective inputs, exploiting statistical assumptions underlying learning algorithms, or leveraging feedback-driven automation~\cite{ballet2019imperceptible,cartella2021adversarial,wawrowski2025adversarial,melo2023adversarial}. Prior work has demonstrated that, under realistic financial constraints such as limited modification budgets, non-IID data, and plausibility requirements, adversarial manipulations can induce misclassification, biased decisions, or long-horizon drift without triggering immediate anomalies~\cite{fursov2021adversarial,goldblum2021adversarial,raff2025adversarial,simonetto2024towards,he2025investigating}. These observations indicate that security threats to financial AI are fundamentally algorithmic and system-level, exploiting learning and inference mechanisms rather than implementation bugs. 

Despite growing attention to cybersecurity and AI risk in finance, existing literature exhibits a structural mismatch with the operational reality of financial AI pipelines. Finance-facing security research predominantly frames AI as a tool for threat detection and cyber defence, emphasising deployment challenges, compliance, and operational assurance~\cite{al2024artificial,javaheri2024cybersecurity,aleksandrova2023survey,mccarthy2022functionality}. While valuable, this perspective treats AI primarily as a defensive capability rather than as an attack surface shaped by learning dynamics. Conversely, the adversarial machine learning (AML) literature has developed rich taxonomies of attack techniques and defenses~\cite{malik2024systematic,vassilev2024adversarial,sadeghi2020system,bountakas2023defense}, but remains largely domain-agnostic, abstracting away constraints that are central in finance, such as market-mediated input channels, cost and feasibility bounded adversarial actions, continuous retraining, and automation-amplified downstream effects~\cite{dyrmishi2025insights,simonetto2024towards,grini2025constrained}. As a result, existing surveys often catalogue attack categories without explaining where and how these attacks manifest within real financial AI pipelines, or why certain mechanisms are especially persistent or damaging in financial settings. 

This gap motivates the need for an integrative analytical framework that is both lifecycle-centric and mechanism-driven. In this survey, we argue that security and robustness in financial AI can only be systematically understood by explicitly connecting adversarial mechanisms to concrete stages of the financial AI lifecycle, and by analysing how algorithmic effects accumulate and persist under continuous operation~\cite{kreuzberger2023machine}. Rather than organising prior works by application domains or attack labels alone, we structure this survey around two coupled axes: (i) a lifecycle-centric view that maps threats to stages of continuously operated financial AI pipelines, where finance-specific constraints are embedded in the end-to-end workflow and shape both attack feasibility and downstream impact; and (ii) an algorithm-/mechanism-centric view that captures the learning and inference computations exploited by adversaries. 

Concretely, we conceptualise continuously operated financial AI as an end-to-end pipeline with three stages: training and model updating, deployment and inference, and operation, monitoring, and feedback~\cite{li2023trustworthy,chen2024security}. In each stage, we characterise the stage-specific attack surfaces and analyse the underlying learning and inference mechanisms that adversaries can exploit, explicitly accounting for how finance-specific constraints embedded in the workflow shape feasibility and downstream impact. The detailed taxonomy, definitions, and the mapping between lifecycle stages and mechanisms are presented in Section~\ref{sec:taxonomy}.

The scope of this survey is deliberately technical. We focus on algorithmic attacks that compromise decision integrity, availability, or trust in continuously operated financial AI pipelines. Adjacent trust dimensions, including governance, compliance, privacy, and reliability, are considered only insofar as they constrain attack feasibility, influence evaluation assumptions, or shape operational impact~\cite{toreini2020relationship}. Broader economic, regulatory, and policy analyses fall outside the scope of this work.

The contributions of this survey are threefold.
\begin{itemize}
    \item We present a lifecycle-centric organisation of security and robustness risks in financial AI pipelines, mapping attack surfaces to stages of continuous operation while embedding finance-specific constraints that shape feasibility and downstream impact.
    \item We develop a mechanism-driven analytical framework that characterises how adversaries exploit learning and inference processes, and explains how algorithmic effects can propagate, accumulate, and persist under continuous deployment, iterative updating, and automated decision execution.
    \item We identify open challenges and under-explored directions for securing financial AI pipelines, including persistent manipulation across updates, robustness of monitoring signals, and emerging operational threats in AI-mediated workflows.
\end{itemize}

Section~\ref{sec:methodology} describes the review methodology and paper selection procedure. Section~\ref{sec:taxonomy} introduces our lifecycle-centric and mechanism-driven taxonomy for financial AI security risks. Sections~\ref{sec:model-training}, \ref{sec:model-deploy}, and \ref{sec:model-monitoring} then analyse security threats across the financial AI lifecycle, covering training and updating, deployment-time inference and decision execution, and operation, monitoring, and feedback. Section~\ref{sec:lit-review} provides a literature comparison, positioning our synthesis against prior surveys and related lines of work. Finally, Section~\ref{sec:conclusion} summarises open challenges and concludes the survey.

\section{Research Methodology}
\label{sec:methodology}

This survey develops an analytical framework to examine security and robustness threats in \emph{continuously operated financial AI pipelines} through lifecycle-centric and mechanism-centric lenses. We use this framing to guide paper search, screening, coding, and taxonomy construction, while recording finance-specific constraints as contextual factors that affect attack feasibility and downstream impact~\cite{page2021prisma}.

\subsection{Scope and Search Strategy}
\label{sec:scope-search}
We focus on \emph{algorithmic} threats that compromise decision integrity, availability, or trust in production-like financial AI systems~\cite{biggio2018wild,liu2022trustworthy}. The covered settings span training and updating, deployment and inference, and operation and monitoring with feedback loops, consistent with the lifecycle framing of this survey~\cite{kumar2020adversarial}. Governance, compliance, privacy, and reliability are considered only when they constrain attack feasibility, influence evaluation assumptions, or shape operational impact~\cite{lecuyer2018certified}.

We conducted a structured search over Google Scholar, ACM Digital Library, IEEE Xplore, SpringerLink, Elsevier ScienceDirect, arXiv, and major security/ML conference proceedings. Query groups combined \emph{finance} terms with \emph{security/robustness} terms and \emph{lifecycle-stage} terms, such as finance/AML + adversarial or poisoning, training/retraining + backdoor or model poisoning, inference/deployment + evasion or black-box, operation/monitoring + prompt injection or deepfake, and modality-specific terms such as time series, tabular, order book, and financial reporting~\cite{ozbayoglu2020deep,jiang2026representation}. We further used backward and forward snowballing to identify closely related and follow-up studies~\cite{wohlin2014guidelines}.

\subsection{Selection Criteria and Screening Procedure}
\label{sec:selection-screening}
We include studies that propose, evaluate, or theoretically analyze attacks and defenses applicable to financial AI tasks; specify threat models, perturbation constraints, or optimization objectives; and provide empirical, mechanistic, or efficiency evidence supporting their claims. We prioritize production-relevant settings such as continuous updating, federated learning, black-box constraints, and operational feedback loops.

We exclude works that focus only on conventional cybersecurity without an algorithmic learning component, use AI purely as a defensive tool without analyzing vulnerabilities of the AI system itself, or provide insufficient technical detail to recover the mechanism, constraints, or evaluation setup. Purely policy, legal, or economic discussions without a material connection to learning or inference vulnerabilities are also excluded.

Screening followed a multi-stage process: title and abstract screening to remove clearly irrelevant papers, followed by full-text screening to verify the presence of an explicit threat or failure model, a concrete attack or defense mechanism, and supporting evaluation evidence. When multiple versions of the same work existed, we prioritized the most complete peer-reviewed version while retaining preprints only when they contained materially new technical details.

\subsection{Data Extraction, Coding, and Taxonomy Construction}
\label{sec:data-coding-taxonomy}
For each retained paper, the data extraction covers: lifecycle stage; attack entry point; exploited mechanism; threat model and budget or feasibility constraints; evaluation metrics, datasets, models, and complexity profile; and the targeted financial task and downstream operational consequence~\cite{goodfellow2014explaining,papernot2016transferability,carlini2017towards,biggio2013evasion}. This process results in a consolidated table that enables consistent cross-task analysis of stealth, persistence, efficiency, and systemic risk. We further develop the Financial AI Security \& Robustness Taxonomy (Figure~\ref{fig:taxonomy}) iteratively by reconciling each paper's operational intervention point with the learning or inference mechanism it exploits, and by checking consistency with the stated threat model and evaluation setting. This procedure keeps the taxonomy grounded in continuously operated financial AI pipelines while supporting cross-task comparison. The resulting category definitions and mappings are presented in Section~\ref{sec:taxonomy}.

\subsection{Quality Control and Limitations}
\label{sec:method-limitations}
To reduce selection bias, we triangulated evidence across multiple venues and checked whether each paper’s claimed financial applicability is consistent with its mechanism, constraints, and evaluation setting. Nevertheless, three limitations remain: emerging areas such as LLM-agent workflow attacks evolve faster than peer review~\cite{greshake2023not}; proprietary financial datasets and deployment settings restrict direct reproducibility~\cite{kapoor2023leakage}; and many defenses are still validated on simplified benchmarks rather than real financial operating conditions. These gaps motivate the open challenges discussed in the final section.

\section{Lifecycle and Mechanism-Driven Taxonomy}
\label{sec:taxonomy}

\begin{figure*}[t]
\centering
\includegraphics[width=.9\textwidth]{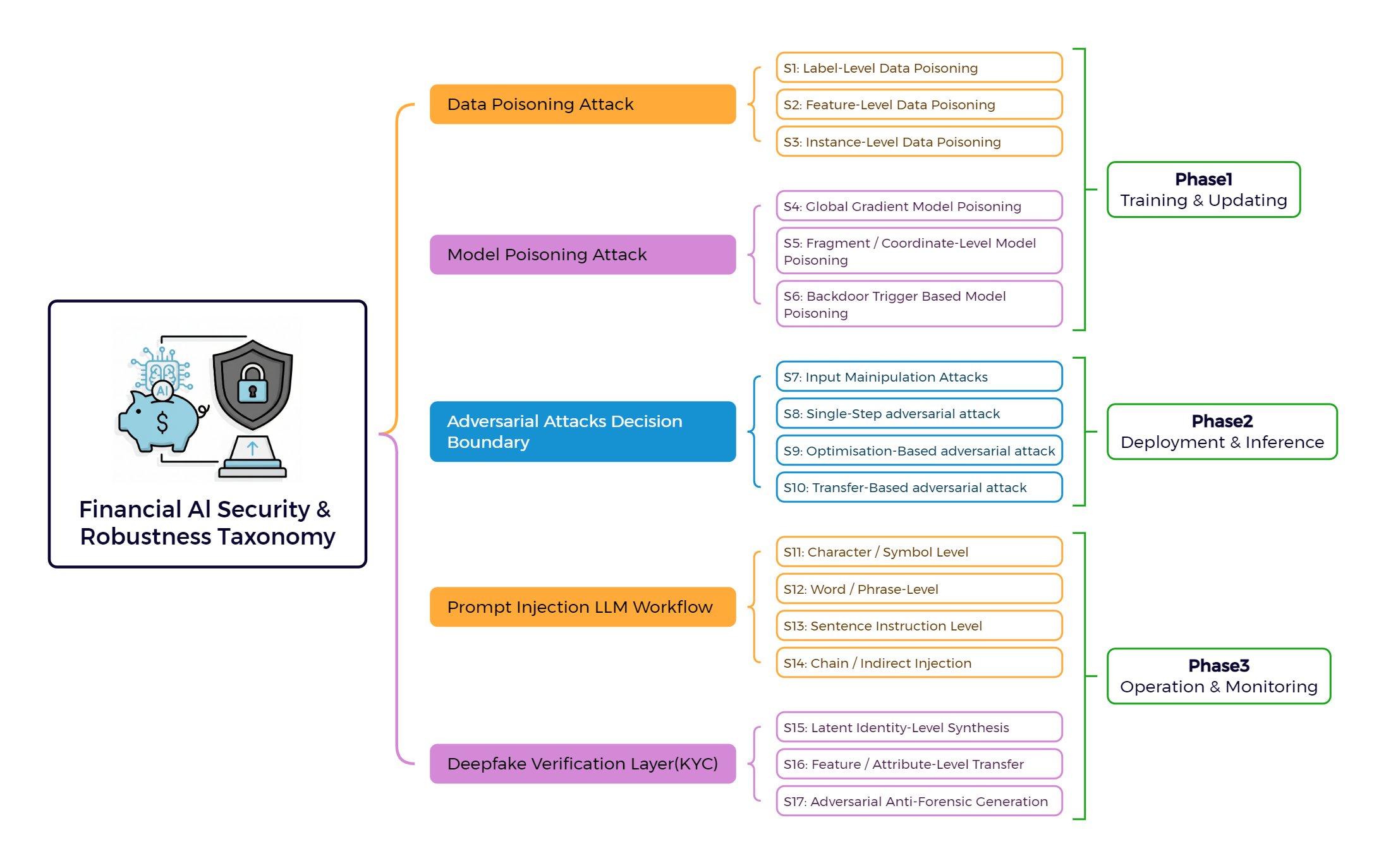}
\caption{
Financial AI Security \& Robustness Taxonomy.
The figure illustrates the proposed lifecycle-centric and mechanism-grounded taxonomy for continuously operated financial AI pipelines. Security and robustness threats are first mapped to three operational phases, namely training \& updating, deployment \& inference, and operation \& monitoring, and are then refined into stage-specific attack classes and subtypes according to the dominant learning or inference vulnerabilities exploited by the adversary.
}
\label{fig:taxonomy}
\end{figure*}

Modern financial AI systems expose a heterogeneous attack surface spanning data pipelines, model optimisation, deployment interfaces, and post-deployment monitoring. Adversarial risks therefore emerge not only from isolated techniques, but from how learning and inference mechanisms interact with structured financial data and continuously operated workflows. To systematically characterise these risks, we propose a unified \textit{Financial AI Security \& Robustness Taxonomy} (Figure~\ref{fig:taxonomy}). The taxonomy is organised along two coupled dimensions. First, we adopt a lifecycle-centric view and partition the end-to-end pipeline into three operational stages: \emph{training and updating}, \emph{deployment and inference}, and \emph{operation with monitoring and feedback}. 

These stages encode the finance-specific constraints and trust boundaries inherent to real deployments, which shape both attack feasibility and downstream impact. Second, within each stage, we use a mechanism-centric lens tailored to the dominant learning or inference vulnerabilities of that stage to categorise vulnerabilities by the learning or inference computations an adversary manipulates, yielding mechanism-grounded subtypes that remain comparable across tasks and model classes. Phase 3 captures operational threats that arise after core model deployment and act through workflow-facing or human-interaction interfaces, including AI-mediated workflow manipulation and identity-verification subversion. The resulting stage-wise mapping (Table~\ref{tab:taxonomy-overview}) provides a unifying lens for analysing security and robustness risks across continuously operated financial AI pipelines.

\small
\setlength{\LTcapwidth}{\textwidth}

\begin{longtable}{@{}p{3cm}p{3.5cm}p{9cm}@{}}

\caption{Hierarchical mapping of the proposed Financial AI Security \& Robustness Taxonomy (S1--S17). The taxonomy links attack entry points, attack classes, and representative financial vulnerability contexts across different lifecycle stages.}
\label{tab:taxonomy-overview} \\

\toprule
\textbf{Attack Entry Point} &
\textbf{Attack Class} &
\textbf{Representative Financial Vulnerability Contexts} \\
\midrule
\endfirsthead

\toprule
\textbf{Attack Entry Point} &
\textbf{Attack Class} &
\textbf{Representative Financial Vulnerability Contexts} \\
\midrule
\endhead

\midrule
\multicolumn{3}{r}{\textit{Continued on next page}} \\
\endfoot

\bottomrule
\endlastfoot


\rowcolor{blue!10}   
\multicolumn{3}{l}{\textbf{Phase 1: Training \& Updating}} \\

\multirow{3}{=}{Data Poisoning}

& \textbf{S1: Label-Level Data Poisoning}
& Label manipulation in financial datasets—including profit-driven label flipping in time-series prediction~\cite{gallagher2022investigating} and random label corruption in fraud detection~\cite{kure2025detecting}—distorts trading signals and increases fraud misclassification losses. \\

& \textbf{S2: Feature-Level Data Poisoning}
& Feature perturbations in financial learning pipelines—including bounded AR-feature manipulation in forecasting models~\cite{alfeld2016data} and clustering feature distortion in transaction analysis~\cite{villani2026sonic}—induce systematic forecast bias and degrade financial risk detection. \\

& \textbf{S3: Instance-Level Data Poisoning}
& Injection of adversarial training instances—including camouflaged samples in compliance models~\cite{salem2021get}, trigger-poisoned authentication images~\cite{kutbi2025impact}, and incremental retraining manipulation~\cite{zhan2024effectiveness}—gradually shifts decision boundaries in credit or fraud classification systems. \\

\midrule

\multirow{3}{=}{Model Poisoning}

& \textbf{S4: Global Gradient Model Poisoning}
& Manipulation of aggregated client gradients in federated financial learning systems—via cross-round directional consistency~\cite{xie2025model} or sign-statistics mimicry~\cite{yang2025enhanced}—induces cumulative model drift in credit risk or fraud detection models. \\

& \textbf{S5: Fragment / Coordinate-Level Model Poisoning}
& Coordinate-level gradient perturbations in federated learning—including high-leverage parameter manipulation~\cite{ren2024fmpa} and geometric anchor bypass updates~\cite{Kasyap2024Sine}—enable stealthy degradation of credit risk or AML detection systems. \\

& \textbf{S6: Backdoor Trigger-Based Model Poisoning}
& Trigger-correlated updates in federated financial learning—including clean-label trigger synthesis in vertical FL~\cite{Chen2024Universal,Yang2025Stealthy} and density-adaptive distributed triggers in horizontal FL~\cite{wang2025improved}—implant covert decision rules affecting credit approval or fraud screening. \\


\rowcolor{green!10}  
\multicolumn{3}{l}{\textbf{Phase 2: Deployment \& Inference}} \\

\multirow{4}{=}{Adversarial Attacks on Decision Boundary}

& \textbf{S7: Input Manipulation Attacks}
& Localised input perturbations in financial AI—including timestep edits in time-series forecasting~\cite{fursov2021adversarial}, sparse feature manipulation in fraud detection~\cite{pandey2023improving}, and accounting-consistent reporting adjustments~\cite{raff2025adversarial}—induce prediction flips while preserving statistical plausibility. \\

& \textbf{S8: Single-Step Adversarial Attack}
& Gradient-aligned single-step perturbations (FGSM) applied at inference time exploit local linearity in financial models, causing systematic degradation of forecasting accuracy and cumulative trading losses~\cite{goodfellow2014explaining,gallagher2022investigating}. \\

& \textbf{S9: Optimisation-Based Adversarial Attack}
& Iterative optimisation-based perturbations (e.g., PGD) manipulate structured financial inputs under domain constraints such as liquidity budgets or accounting identities, inducing worst-case prediction errors in trading or regulatory models~\cite{madry2017towards,goldblum2021adversarial}. \\

& \textbf{S10: Transfer-Based Adversarial Attack}
& Transferable perturbations generated on surrogate models exploit cross-model decision-boundary alignment, enabling black-box manipulation of proprietary financial AI systems such as trading agents or fraud detection ensembles~\cite{papernot2016transferability,moosavi2017universal,piazza2021adversarial,fok2025foe}. \\


\rowcolor{orange!10} 
\multicolumn{3}{l}{\textbf{Phase 3: Operation \& Monitoring}} \\

\multirow{4}{=}{Prompt Injection (LLM Workflow)}

& \textbf{S11: Character / Symbol-Level}
& Encoding-level perturbations in LLM input streams—such as zero-width characters, homoglyph substitution, or Unicode tag smuggling—exploit tokenization and normalization inconsistencies to bypass safety filters and induce unintended disclosure or privileged actions~\cite{sang2024evaluating,hackett2025bypassing}. \\

& \textbf{S12: Word / Phrase-Level}
& Token-level semantic or structural manipulations—including latent-space semantic optimisation~\cite{liu2025semantics} and homoglyph-based visual deception in financial news streams~\cite{rizvani2026adversarial}—can distort event interpretation in financial prediction or trading pipelines. \\

& \textbf{S13: Sentence / Instruction-Level}
& Injection of adversarial instructions within natural-language task templates—including compositional instruction attacks~\cite{jiang2023prompt} and gradient-optimised prompt leaking~\cite{hui2024pleak}—can override system policies and manipulate automated compliance or advisory decisions. \\

& \textbf{S14: Chain / Indirect Injection}
& Malicious instructions embedded in external content sources (e.g., RAG documents or API outputs) can hijack tool-using financial agents by reinterpreting retrieved data as executable instructions, enabling data exfiltration or transaction manipulation~\cite{alizadeh2025simple,liu2025autohijacker}. \\

\midrule

\multirow{3}{=}{Deepfake Verification Layer (KYC)}

& \textbf{S15: Latent Identity-Level Synthesis}
& Generative identity synthesis in latent space—including identity-controlled diffusion generation~\cite{papantoniou2024arc2face} and stochastic master-face search~\cite{nguyen2022master}—can produce synthetic identities that bypass biometric similarity thresholds in financial onboarding or watchlist screening. \\

& \textbf{S16: Feature / Attribute-Level Transfer}
& Identity–attribute disentanglement in face-swapping models—including diffusion-based identity transfer~\cite{zhao2023diffswap} and automated liveness-response synthesis~\cite{li2022seeing}—enables real-time impersonation that bypasses facial verification in remote KYC and account recovery workflows. \\

& \textbf{S17: Adversarial Anti-Forensic Generation}
& Adversarial multimedia generation—including physical optical projection attacks on 3D sensors~\cite{li2023physical} and query-efficient decision-based evasion against black-box APIs~\cite{dong2019efficient}—can circumvent biometric detection systems and compromise financial authentication pipelines. \\

\end{longtable}


\section{Model Training Phase}
\label{sec:model-training}

The model training and updating phase constitutes the foundational stage of the financial AI lifecycle, where algorithms ingest historical financial data, ranging from transaction logs to credit profiles, to construct the core decision-making logic. In this stage, the primary objective is optimization: algorithms iteratively adjust internal parameters to minimize statistical errors and capture underlying data distributions. However, this purely mathematical optimization introduces a structural vulnerability. The training mechanism inherently assumes that the provided inputs—whether centralized datasets or distributed parameter updates—are accurate representations of the legitimate financial environment and are free from malicious intent~\cite{steinhardt2017certified,bagdasaryan2020backdoor}. Because the learning process lacks contextual understanding and blindly trusts its inputs, any manipulation at this stage becomes deeply embedded in the model's fundamental ``worldview.'' Consequently, adversarial interventions during training establish a compromised baseline. These embedded flaws seamlessly propagate downstream, where subsequent automated execution and feedback loops amplify them into systemic financial risks. This structural reliance on input integrity gives rise to two primary vectors of compromise: manipulating the underlying historical records, or hijacking the decentralized optimization process.

\subsection{Data Poisoning}
\label{subsec:data-poisoning}

Data poisoning occurs when adversaries contaminate the historical datasets used to train or retrain financial models. By strategically altering specific input features or target labels, attackers embed malicious statistical associations into the model's logic. Rather than directly attacking the algorithm, this mechanism manipulates the environment from which the algorithm learns. 

In the financial sector, these localized data modifications trigger severe operational disruptions across various domains. In credit scoring systems, attackers might invert default labels on specific historical profiles, inducing the model to systematically approve high-risk loans or deny legitimate borrowers\cite{patil2024analyzing}. Within anti-money laundering (AML) and fraud detection pipelines, the insertion of carefully crafted benign-looking fraudulent transactions—or vice versa—skews the model's decision boundaries\cite{paladini2023fraud}. This directly degrades detection efficacy, leading to increased false negatives that facilitate illicit fund transfers, or overwhelming compliance teams with operational false positives. Furthermore, in algorithmic trading, polluting high-frequency data streams or feature engineering outputs can force execution engines to initiate erroneous trades at critical market junctures, resulting in immediate capital loss\cite{faghan2020adversarial}. Because these poisoned associations are deeply encoded into the model's weights, they are notoriously difficult to isolate and remove, ensuring that the resulting compliance failures and financial losses persist throughout the deployment lifecycle.

\subsubsection{Label-Level Poisoning}
\label{subsubsec:label-level-poisoning}

Label-level poisoning changes the class labels in training data without modifying input features, which confuses decision boundaries during learning and induces persistent errors at test time. In finance, this threat can be injected into credit scoring, transaction prediction and anti--money--laundering pipelines, creating stealthy and hard--to--repair misclassification and elevating systemic risk.


Gallagher et al.\ (2022) study label-level poisoning for financial time-series prediction using public datasets (stock prices, futures returns) with 1D--CNN forecasters\cite{kiranyaz20211d}, proposing \emph{Profit-Based Targeted Label Injection}~\cite{gallagher2022investigating}. A clean baseline is trained on rolling windows; each windowed sample $x_i$ is then backtested to obtain its profitability score $P(x_i)$, samples are sorted from least to most profitable, and the worst $k=\lceil n\%\cdot N\rceil$ have their buy/sell labels swapped before the model is retrained on the merged data. Time complexity is $O(N\log N)$ ($O(N)$ backtesting plus $O(N\log N)$ sorting) with $O(N)$ space for profitability scores. Experiments show that flipping only $10\%$ of samples increases losses by more than $130\%$, delivering maximal financial damage under a very small poisoning budget. However, this computationally expensive $O(N \log N)$ strategy relies on deterministic sorting of validation signals, raising the question of whether similar degradation is achievable through cheaper $O(N)$ stochastic techniques that do not require explicit sample importance ranking.


Validating the efficacy of stochastic corruption over deterministic optimization, Kure et al.\ (2025) investigate \emph{Random Label Flipping} and \emph{Image Replacement} on both CIFAR--10 and insurance fraud detection datasets~\cite{kure2025detecting}. Rather than searching for optimal samples, the adversary selects a fixed proportion $p$ of cross--class pairs for random label swapping, reducing time complexity to linear $O(N)$ by bypassing the sorting phase. Despite lacking targeted selection, the attack reduces ResNet--50 top--1 accuracy on CIFAR--10 from $92.3\%$ to $65.1\%$. Crucially, in the financial domain (insurance claims), flipping just $5\%$ of labels causes fraud detection recall to plummet from $88\%$ to $47\%$, confirming that probabilistic attacks favouring distribution--wide statistical corruption over individual sample importance can induce severe misclassification in financial pipelines without domain--specific backtesting or gradient oracles.

The two strategies illustrate a fundamental algorithmic trade-off between \emph{optimization-based precision} and \emph{stochastic scalability}. Gallagher et al.\ employ a greedy heuristic (profit sorting) to identify high-leverage samples, achieving maximal damage with minimal data modification but at a superlinear computational cost ($O(N \log N)$). In contrast, Kure et al.\ demonstrate that probabilistic attacks can achieve comparable systemic failures with linear complexity ($O(N)$). While optimization yields stealthiness by targeting fewer samples, stochastic methods offer broader applicability in black--box financial scenarios where gradient or profitability oracles are unavailable, suggesting that defence mechanisms must be robust against both precise surgical strikes and broad statistical noise.

\subsubsection{Feature-Level Poisoning}
\label{subsubsec:feature-level-poisoning}

Feature-level poisoning injects fine-grained and targeted perturbations into input features while keeping labels unchanged, which skews the learner’s attribution to key predictors during training and yields persistent errors at test time. In financial pipelines, adversaries can hide noise inside sensitive dimensions such as transaction frequency, account balance and market microstructure so that downstream decisions drift toward misapproval in lending, under-reporting in fraud detection or unintended deviations in algorithmic trading.

Alfeld et al.\ (2016) propose a poisoning framework for time-series forecasting under a fixed and known linear AR($d$) model with recursive prediction~\cite{alfeld2016data}. The attacker perturbs only a subset of entries in the initial input vector with bounded magnitude $\delta$ to minimise the weighted discrepancy between the poisoned forecast path and a target path. Constraints such as per-step and total budgets, nonnegativity, sparsity, and quadratic effort are formulated as convex or quadratic programs, yielding time complexity $O(k^3)$ and space complexity $O(k^2)$, where $k$ is the number of editable feature dimensions. The key mechanism is that recursive prediction amplifies small early-step perturbations along the time axis, allowing limited-magnitude edits to accumulate into systematic bias over $h$ future steps. Experiments on synthetic sequences and real natural-gas prices show that a total budget as low as $10\%$ induces pronounced spikes or dips at designated times while consistently outperforming greedy and random baselines. Since the method targets the prevalent linear AR recursion in financial time series and is validated on real commodity prices with feasible budgets, it is representative and practically threatening for finance, and the same mechanism transfers to volatility and volume forecasting, liquidity prediction, and temporal feature estimation for credit scoring when only upstream inputs are editable; however, it relies on convex loss surfaces and thus lacks flexibility for attacking non-differentiable metrics in discrete grouping tasks.

Shifting from deterministic solvers to evolutionary heuristics to tackle such non-convex complexities, Villani et al.\ (2024) present \emph{Sonic}, a feature-level poisoning method that disrupts clustering structure~\cite{villani2026sonic}. Sonic uses an incremental density-based proxy, FISHDBC, to avoid rerunning the full target clustering each iteration, for example DBSCAN\cite{ester1996density} or HDBSCAN\cite{campello2013density}, and performs genetic search on a controlled subset of samples using adjusted mutual information as the fitness signal. By updating incrementally on a subset of size $m\ll n$, computation is reduced from roughly $O(n\log n)$ to $O(m\log m)$ in time and from $O(n)$ to $O(m)$ in space. On image and text benchmarks, with poisoned-sample ratios of $1\%\text{--}20\%$ and perturbation bounds of $5\%\text{--}60\%$, Sonic consistently lowers AMI and, for comparable disruption, runs orders of magnitude faster than directly attacking the target clustering. Although evaluated outside finance, its ability to alter density or hierarchy with few poisoned samples and small feature edits transfers naturally to unsupervised financial tasks such as anti--money--laundering cell merging or splitting, abnormal transaction pattern mining, and customer or merchant risk clustering.

Both studies operate at the feature level yet rely on different carriers and optimisation schemes. Alfeld et al.\ provide analytically solvable convex or quadratic programs for precise perturbations under known linear AR($d$), which yields a $O(k^3)$ time profile suitable for structured forecasting tasks. Sonic deploys genetic search with an incremental density proxy to avoid full reruns, achieves near-global structure disruption on a subset with $m\ll n$ and runs in $O(m\log m)$, which is amenable to black-box unsupervised settings. The AR-based route poses higher risk to indicators with strong temporal recursion such as price, volatility and volume, whereas Sonic is more damaging for network-like tasks with unknown structure such as suspicious-account community discovery and merchant risk clustering. Neither work provides effective detection or defence, and open questions remain on cross-model generalization, co-evolution with defences and long-run adaptation in dynamic financial systems. This contrast clarifies implementation routes and efficiency bottlenecks for feature-level poisoning and supports defence prioritisation in finance; our survey systematises task–attack fit and risk gradients where prior literature lacked cross-algorithm transfer analysis.

\subsubsection{Instance-Level Poisoning}
\label{subsubsec:instance-level-poisoning}

Instance-level poisoning injects a small number of malicious instances—either with specific trigger patterns or with corrupted labels—so that the model learns a backdoor or bias during training and later outputs attacker-preferred results whenever inputs contain the same trigger. In financial applications, such poisoned instances can be blended into data pipelines and silently steer decisions once the trigger appears.

Salem et al.\ (2022) propose a training-time model hijacking scheme that embeds a hidden \emph{auxiliary task} while preserving original-task utility~\cite{salem2021get}. A \emph{Camouflager} encoder--decoder transforms auxiliary-task samples into \emph{camouflaged} instances that resemble the original distribution but map to attacker-chosen labels, with two variants: \emph{Chameleon}, which balances appearance and semantics, and \emph{Adverse Chameleon}, which adds adversarial constraints for stricter settings. Overheads grow linearly with the camouflager size and the number of poisoned instances. Experiments show high hijacking success with minimal utility loss---e.g., CIFAR-10/MNIST reaches about $99\%$ attack success with only a $0.5\%$ accuracy drop, while CIFAR-10/CelebA\cite{liu2015deep} achieves about $73.7\%$ with a $3.8\%$ drop. The method transfers to image-based financial compliance workflows; however, such global repurposing requires complex digital camouflage, and specific identity-verification scenarios often face simpler physically realisable threats that bypass detection without optimising the entire task manifold.

Kutbi (2025) evaluates training-time trigger-based backdoor attacks for face classification by injecting small trigger patterns into selected training images with a common target label~\cite{kutbi2025impact}, examining how trigger location, size, and shape trade off stealthiness and success. Retraining cost scales linearly with network size and poisoned samples, while inference adds no overhead. Even very small poisoning rates yield attack success approaching $100\%$ with clean accuracy nearly unchanged, and a small fixed-location trigger provides a strong balance between stealth and efficiency. The technique transfers directly to face-driven financial authentication such as face payment and account login; however, this static offline setting leaves open how poisoning exploits continuous update mechanisms in modern financial systems.

Zhan et al.\ (2024) examine black-box instance-level poisoning in which an attacker injects a few malicious instances each round to gradually shift the decision boundary~\cite{zhan2024effectiveness}. The key mechanism exploits the history dependence of online fine-tuning, so small per-round disturbances accumulate into a persistent bias over time. Training cost grows approximately linearly with the number of poisoning rounds and injected samples, while memory overhead remains close to the baseline. On synthetic 2D data, the misclassification rate rises from $0\%$ at round 1 to $0.21\%$ at round 5, $0.73\%$ at round 10, and $30\%$ at round 15. The method transfers to financial online retraining pipelines such as credit-default and fraud models, where stealthy long-horizon injection can shift thresholds while headline accuracy appears stable; however, simple incremental injection lacks the capacity to model the complex sparse distributions typical of user--service interactions.

Chen et al.\ (2025) propose instance-level poisoning for a QoS-aware cloud API recommender by synthesising realistic user interactions to boost target API rankings~\cite{chen2025latent}. Their \emph{compress-then-diffuse} mechanism first uses autoencoders to compress the interaction matrix into latent representations and then trains a diffusion model with an adversarial attack term to generate forged interactions that remain close to the genuine distribution. The attacker modifies only newly created interactions without touching existing data, improving stealth under a plausibility constraint. On WS-DREAM\cite{zheng2012investigating} with a $10\%$ budget and attack strength 20, the attack improves HR@50 of PMF by about $127\%$. The approach transfers to financial recommendation and profiling scenarios such as wealth-product recommendation and merchant ranking, where injecting realistic \emph{pseudo-users} can shift rankings with minimal headline impact, creating compliance and risk-control concerns.

These studies collectively illustrate the spectrum of instance-level poisoning, ranging from static backdoor injection to dynamic boundary shifting and generative masquerading. Salem et al.\ and Kutbi focus on fixed training sets but target different vulnerabilities: the former employs complex encoder–decoder optimisation for global task hijacking suitable for compliance evasion, while the latter utilises simple, physically realisable triggers for efficient authentication bypass. In contrast, Zhan et al.\ and Chen et al.\ address dynamic and distributional complexities; Zhan et al.\ leverage history-dependent incremental updates to induce gradual drift in credit or fraud models with linear complexity, whereas Chen et al.\ deploy computationally intensive diffusion processes ($O(\text{steps} \times \text{dim})$) to synthesise high-fidelity interactions for recommender systems. The static backdoor approaches pose immediate threats to biometric and identity verification workflows, while the dynamic and generative methods imperil adaptive systems like online risk scoring and personalised financial services where distribution shifts are subtle. Similar to feature-level attacks, robust defences for these instance-based manipulations remain under-explored, particularly regarding the trade-off between inspection cost and detection latency in high-throughput financial pipelines. Our survey clarifies these operational boundaries, mapping distinct poisoning modalities to their corresponding financial risk surfaces—from login security to long-term credit model stability—where previous works treated them in isolation.

\subsubsection{Comparative Analysis of Data Poisoning Subtypes}

Table~\ref{tab:data-poisoning-summary} systematically compares the three data poisoning subtypes across four critical dimensions: algorithmic strategy, complexity profile, stealth, and systemic risk priority. Across these variations, injection prerequisites and stealth mechanisms directly dictate the magnitude of financial disruption. Instance-level interventions present the highest systemic risk priority due to their negligible inference overhead, physical realizability, and capacity to induce long-horizon algorithmic drift in automated approvals with extremely low injection budgets. Conversely, feature-level and label-level attacks, while highly damaging to specific quantitative or compliance tasks, demand stricter prerequisites regarding upstream data access. They often rely on more mathematically rigid optimization routines—ranging from $O(N \log N)$ sorting to $O(k^3)$ convex solvers—and consequently exhibit tighter bounds on cross-algorithm transferability and operational stealth.

Rather than representing a simple chronological progression, the literature within each subclass is deliberately organized to illustrate distinct methodological trade-offs dictated by varying financial deployment constraints. At the label level, the primary contrast lies in optimization complexity, balancing high-cost deterministic sorting against highly scalable stochastic flipping to induce cumulative financial loss. At the feature level, the strategic focus shifts from white-box analytical solvers for structured commodity pricing to black-box heuristic searches that stealthily disrupt unsupervised clustering topologies without requiring full structural reruns. Finally, at the instance level, the operational threat diverges from static global backdoors designed for verification bypass toward dynamic boundary drift and generative forging that actively exploit continuous retraining pipelines. By mapping these algorithmic contrasts directly to their financial targets, it becomes evident that the most persistent systemic risks arise not from arbitrary data corruption, but from highly targeted interventions tailored to the specific optimization and deployment vulnerabilities of financial pipelines.

\begin{table}[htbp]
\small
\centering
\caption{Comparison of data poisoning subtypes in the training phase, highlighting methodological trade-offs, deployment prerequisites, persistence mechanisms, and their implications for financial tasks and downstream operational risk.}
\label{tab:data-poisoning-summary}
\renewcommand{\arraystretch}{1.3}
\setlength{\tabcolsep}{6pt}
\begin{tabularx}{\textwidth}{l
                            >{\raggedright\arraybackslash}X
                            >{\raggedright\arraybackslash}X
                            >{\raggedright\arraybackslash}X}
\toprule
\textbf{Dimension} &
\textbf{a1 Label-Level Poisoning} &
\textbf{a2 Feature-Level Poisoning} &
\textbf{a3 Instance-Level Poisoning} \\
\midrule
\multicolumn{4}{c}{\textit{Algorithmic \& Methodological Contrast}} \\
\midrule
\textbf{Algorithmic Strategy} &
Greedy profit sorting~\cite{gallagher2022investigating} vs. uniform stochastic flipping~\cite{kure2025detecting}. &
Analytical convex programming~\cite{alfeld2016data} vs. heuristic genetic search on proxy structures~\cite{villani2026sonic}. &
Static global/local backdoors~\cite{salem2021get,kutbi2025impact} vs. incremental boundary drift and generative forging~\cite{zhan2024effectiveness,chen2025latent}. \\

\textbf{Complexity Profile} &
$O(N \log N)$ (deterministic) to linear $O(N)$ (stochastic). &
$O(k^3)$ (solvers) to $O(m \log m)$ on controlled proxy subsets. &
Negligible inference overhead; training scales with diffusion steps or multi-round injection. \\

\textbf{Injection Prerequisites} &
Reachable targets in historical data; outsourced or noisy annotation channels. &
Modifiable input features upstream; accessible data engineering or aggregation outputs. &
Ability to inject a small number of physical or digital instances during continuous retraining. \\

\textbf{Stealth \& Persistence} &
Medium stealth; amplified primarily via temporal recursion in autoregressive setups. &
High stealth; structural disruption remains stable across different clustering algorithms. &
Very high stealth; triggers are physically realizable and biases survive multi-round updates. \\

\midrule
\multicolumn{4}{c}{\textit{Financial Deployment \& Systemic Impact}} \\
\midrule
\textbf{Target Financial Task} &
Time-series forecasting~\cite{gallagher2022investigating} and static insurance fraud records~\cite{kure2025detecting}. &
Structured commodity pricing~\cite{alfeld2016data} and unsupervised risk clustering (e.g., AML)~\cite{villani2026sonic}. &
Biometric verification (eKYC)~\cite{kutbi2025impact}, dynamic credit scoring~\cite{zhan2024effectiveness}, and recommendation~\cite{chen2025latent}. \\

\textbf{Downstream Consequence} &
Unintended spikes/dips in quantitative trading; severe recall degradation in compliance alerts. &
Misguided liquidity forecasting; splitting or merging of suspicious account communities. &
Silent bypass of identity workflows; long-horizon algorithmic drift in automated approvals. \\

\textbf{Systemic Risk Priority} &
Medium &
High &
Very High \\
\bottomrule
\end{tabularx}
\end{table}


\subsection{Model Poisoning}
\label{subsec:model-poisoning-fl}

While data poisoning manipulates the underlying historical records, model poisoning directly compromises the mathematical optimization trajectory. This threat is predominantly realized in distributed architectures such as Federated Learning (FL). In an FL paradigm, multiple financial institutions collaborate to train a global model—typically for fraud detection, anti-money laundering (AML), or credit scoring—without ever exchanging sensitive customer data\cite{faghan2020adversarial}. Instead of pooling raw information in a central repository, participants compute model updates (such as gradients or weights) locally using their private datasets, and transmit only these mathematical summaries to a central server for aggregation. 

This decentralized architecture inherently incentivizes model poisoning because it creates a fundamental structural blind spot: data availability without data visibility. To comply with strict privacy regulations and cross-institutional confidentiality requirements, the central orchestrating server is explicitly prohibited from inspecting the local data that generated the updates. Consequently, the server is forced to rely entirely on statistical aggregation protocols (e.g., Krum, Trimmed Mean) to update the global model\cite{blanchard2017machine}. Adversaries systematically exploit this opaque trust boundary. Because the server cannot verify the underlying data integrity, a malicious participant can carefully craft poisoned parameter updates that mathematically bypass statistical anomaly detection, directly injecting malicious logic or backdoors into the shared global model\cite{bagdasaryan2020backdoor}.

In practice, the operational consequences of exploiting this structural vulnerability are severe. A compromised participant in an inter-bank network could submit poisoned gradients to artificially lower the recall rate for specific transaction typologies, effectively engineering a ``safe passage'' for syndicated money laundering. Similarly, in federated credit risk assessment, malicious updates can subtly bias the global model to automatically trigger loan approvals or manipulate risk limits for targeted profiles\cite{naseri2024badvfl}. By attacking the aggregation mechanism itself, adversaries cause direct financial exposure and undermine the trust foundation of the entire financial consortium, all while maintaining complete isolation from the central dataset.

\subsubsection{Global Gradient Poisoning}
\label{subsubsec:global-gradient-poisoning}

In centralised or federated learning, the adversary directly tampers with \emph{round-level} gradient updates (direction, magnitude, or summary statistics) so that the aggregated global model converges toward collapse. Typical designs flip signs, rescale magnitudes, or synthesize pseudo-gradients aligned with an adversarial target, aiming to push the model off a safe trajectory \emph{before} aggregation filters can intervene.

Xie et al.\ (2024) exploit the vulnerability of aggregation rules to \emph{cross-round consistency} rather than single-round magnitude with \emph{PoisonedFL}~\cite{xie2025model}. Targeting standard FL systems where the attacker controls a subset of clients without benign data knowledge, the method optimises the total aggregated update along a fixed random sign vector $\mathbf{s}$, converting weak per-round deviations into significant cumulative drift through strict directional consistency. A dynamic magnitude schedule evades detection by adjusting the scalar $\lambda^t$ via a hypothesis test on whether the previous round's global update followed $\mathbf{s}$, requiring only $O(d)$ per-round complexity. Results on five benchmarks show PoisonedFL degrades models to random-guess capability, bypassing Trimmed Mean\cite{yin2018byzantine} and FLTrust\cite{cao2020fltrust}. However, rigidly enforcing a fixed direction distorts the sign statistics of malicious updates, creating a statistical footprint detectable by advanced defenses.

To overcome this detectability, Yang et al.\ (2025) introduce \emph{ScaleSign}~\cite{yang2025enhanced}, a \emph{statistical mimicry} mechanism tailored to defeat sign-statistics-based defenses such as SignGuard. ScaleSign constructs malicious updates via two coupled components: a \emph{Scaling Attack} adjusting cosine similarity to match benign updates, and a \emph{Sign Modification Component} forcing the gradient's sign distribution (positive, negative, zero counts) to align with a reference benign statistic $(\bar{P}, \bar{Z}, \bar{N})$. This transforms the problem from pure damage maximisation to constrained optimisation where statistical indistinguishability is the primary constraint. Formal analysis confirms that ScaleSign maintains high cosine similarity while preserving toxicity, achieving $98.23\%$ attack success against SignGuard on CIFAR-10---representing an evolution from brute-force directional consistency to evasion-focused stealth.

The evolution from \emph{PoisonedFL} to \emph{ScaleSign} illustrates a shift in the threat landscape from \emph{accumulated magnitude} to \emph{statistical alignment}. \emph{PoisonedFL} demonstrates that simple cross-round consistency allows attackers to penetrate baseline robust aggregators (like Krum or Median) by slowly dragging the model parameters in a unified direction. In contrast, \emph{ScaleSign} addresses the counter-measures to such drag (specifically SignGuard) by ensuring the \emph{distributional features} of the malicious signal mirror honest participants. For financial deployment, this distinction dictates specific monitoring strategies. High-frequency trading or real-time fraud detection models, which aggregate updates rapidly, are vulnerable to the accumulated drift of \emph{PoisonedFL}; defenses here must monitor \emph{cross-round directional autocorrelation} rather than just single-round outliers. Conversely, collaborative compliance models involving multiple institutions are susceptible to \emph{ScaleSign}, as attackers can hide behind the valid statistical profiles of compromised nodes. Security audits in these contexts should not rely solely on cosine similarity or sign checks but must implement \emph{spectral analysis} (like MSGuard proposed in~\cite{yang2025enhanced}) or perform periodic ``replay checks'' on local data to verify that gradient statistics genuinely match the underlying data distribution.

\subsubsection{Fragment/Coordinate-Level Poisoning}
\label{subsubsec:fragment-coordinate-poisoning}

The adversary perturbs only a \emph{small subset of parameters or coordinates} (e.g., selected layers, channels, low-rank subspaces, or coordinate blocks), creating a sparse ``phantom'' that evades robust aggregation and anomaly checks. The effect is persistent misclassification on \emph{specific classes/regions} while headline accuracy barely moves; stealth relies on \emph{matching gradient statistics} (mean/variance/norm) and using small but directionally consistent perturbations under $L_0$/$L_\infty$ budgets.

Yang et al.\ (2025) perform coordinate-level poisoning in FL using \emph{ScaleSign}~\cite{yang2025enhanced}, independently controlling direction and sign distribution per coordinate: scaling toward the benign direction to raise cosine similarity, then matching sign ratios so the malicious vector passes the aggregator's acceptance window while pushing global drift. Per-round cost is $O(d \log d)$ time and $O(d)$ memory, totalling $O(T \cdot m \cdot d \log d)$. Against SignGuard\cite{xu2022byzantine}, trimmed-mean, and distance-thresholding, the attack achieves near $100\%$ success in certain settings with $\approx41.78\%$ multi-defense improvement over baselines. However, the $O(d \log d)$ sorting dependency limits scalability, prompting a shift toward sparse injection strategies.

To circumvent this burden, Ren et al.\ (2024) deliver \emph{FMPA}~\cite{ren2024fmpa}, concentrating manipulation on a few high-leverage coordinates ($k \ll d$, often the final layer) with bounded directional adjustments that keep global statistics within normal ranges to bypass median/trimmed-mean/Krum\cite{blanchard2017machine}. Extra crafting per round is only $O(k)$; total cost accumulates linearly over $T$ and $m$. Error increases persist across five robust aggregators with high stealth, though such sparse attacks remain vulnerable to geometric defenses enforcing directional consistency with a trusted anchor.

Addressing this vulnerability, Kasyap and Tripathy (2024) show with \emph{Sine}~\cite{Kasyap2024Sine} that similarity alone cannot stop local model poisoning, decomposing updates into per-coordinate sign and magnitude to maintain high cosine to a trusted anchor while maximising deviation from the benign mean---evading FLTrust and trimmed-mean at $O(d)$ per-round cost. Results show stable penetration of cosine-only defenses, while a paired defense (FLTC) reduces residual impact to $\approx2\%\text{--}4\%$. In finance consortia with prevalent cosine-to-anchor checks and non-IID drift, this bypass transfers to AML risk modelling and multi-bank credit scoring, arguing for \emph{multi-metric} consistency over single-threshold auditing.

Fragment/coordinate attacks concentrate manipulation on a few high-leverage dimensions and disguise it by alignment (direction, sign ratios, anchor similarity), keeping each update within acceptable statistics while steadily weakening the model—especially hard to detect under non-IID conditions. The synthesis of these works reveals a clear optimization trajectory: from computationally expensive global sorting ($O(d \log d)$) to efficient fragment-level targeting ($O(k)$) and finally to geometric anchor bypass. For financial systems, this implies that defenders cannot rely on a single ``silver bullet'' metric. Specifically, consortium credit profiling and AML models should move beyond simple cosine-similarity checks, which are demonstrably bypassable by \emph{Sine}-like attacks. Instead, a robust defense posture requires monitoring \emph{cross-round sign-distribution drift} (to catch ScaleSign) and \emph{layer-wise residual energy} (to catch FMPA), ensuring that even sparse, high-leverage perturbations are flagged before they can degrade the global decision boundary.

\subsubsection{Backdoor/Trigger-Based Poisoning}
\label{subsubsec:backdoor-trigger-poisoning}

This class embeds a \emph{function-level} malicious behavior into the global model during training while keeping the main-task utility nearly intact. Unlike global-gradient drift, the backdoor forces a \emph{targeted verdict} when a trigger pattern appears. The evolution of these attacks focuses on overcoming two primary constraints: label inaccessibility in Vertical FL (VFL) and robust aggregation defenses in Horizontal FL (HFL).

In VFL scenarios typical of financial credit scoring, the passive party often lacks label access. Chen et al.\ (2024) introduce a \emph{clean-label} universal backdoor attack (UAB) exploiting the ``class skew'' inherent in binary classification~\cite{Chen2024Universal}. The method infers pseudo-labels by analysing active-party gradient norms ($||g||_2$), leveraging the observation that minority-class gradients exhibit distinct magnitudes, then alternates between \emph{universal-trigger synthesis} and \emph{clean-label injection} via bilevel optimisation. On LendingClub and Zhongyuan, ASR exceeds $90\%$ while maintaining main-task AUC. However, reliance on intrinsic gradient statistics limits precision when class distributions are balanced or gradient distinctions are absent.

To address these precision limitations, Yang et al.\ (2025) propose a semantic alignment framework incorporating auxiliary data to refine trigger synthesis in VFL~\cite{Yang2025Stealthy}. A label inference attack trained on a small auxiliary labelled set builds a local reconstruction model; clustering and neighbour assignment then maps local samples to semantically nearest target-class embeddings, with triggers embedded to coercively align source samples with these clusters. On CIFAR-10 and CINIC-10, ASR exceeds $80\%$ with only $2\%\text{--}6\%$ main-task degradation. While these VFL methods focus on inferring missing labels, HFL attacks face a fundamentally different barrier: robust aggregation protocols filtering anomalous updates.

Shifting to the HFL setting, Wang et al.\ (2025) introduce a density-adaptive strategy evading robust aggregation (RFA\cite{pillutla2022robust}, Krum) by decomposing the global trigger into sub-triggers aligned with high-density edge regions of client data via Canny edge detection~\cite{wang2025improved}. A \emph{Projected Gradient Updating} mechanism dynamically constrains malicious updates within the legitimate statistical boundary defined by the global model, ensuring poisoned gradients mimic natural patterns. On COCO\cite{lin2014microsoft}, ASR reaches $91.68\%$ while outperforming baseline DBA in main-task accuracy by over $8\%$, validating the coupling of distribution-aware poisoning with gradient constraints.

Backdoor/trigger poisoning \emph{rewrites functionality} so that triggers enforce target outputs while headline utility remains stable. Two FL topologies and two evasion paths are covered: VFL clean-label \emph{universal triggers} (gradients/neighbour similarity drive pseudo-labelling and trigger updates with labels hidden), VFL \emph{stealth alignment} (pull source embeddings toward target clusters for post-aggregation shifts), and HFL \emph{density-adaptive distributed triggers} with \emph{projected gradients} to remain within aggregation tolerances. For financial FL, VFL is common in cross-institution credit/fraud with label centralization; thus label-invisible trigger training and embedding-neighborhood pulling can implant stable backdoors without violating protocol. HFL is common in multi-bank or multi-channel risk learning; density-split sub-triggers and gradient projection preserve stealth under non-IID data and robust aggregation. Evaluation and governance should differentiate these risk surfaces: prioritize \emph{trigger hit rate} across rounds, \emph{stable clean-utility bandwidth}, and \emph{post-attack residual effect}; add \emph{class-conditional embedding drift} and \emph{trigger probes} to routine audits so ``near-constant utility yet rewritten function'' is surfaced early in joint modelling pipelines.

\subsubsection{Comparative Analysis of Model Poisoning Subtypes}

Table~\ref{tab:model-poisoning-summary} systematically contrasts the three model poisoning subtypes across algorithmic strategy, stealth mechanisms, and systemic risk priority. The financial impact of these attacks is directly correlated with their ability to exploit the specific transparency gaps in distributed learning. Global gradient poisoning presents a medium risk, primarily threatening the convergence of high-frequency trading or real-time systems through accumulated drift, though it remains susceptible to advanced spectral auditing. Fragment and coordinate-level attacks elevate the risk by exploiting the non-IID nature of financial consortia (e.g., AML networks), where sparse, high-leverage perturbations can be masked as legitimate institutional heterogeneity. Backdoor attacks represent the highest systemic priority; by rewriting the model's functional logic to include hidden "safe passages" for credit or laundering fraud, they create long-lived vulnerabilities that persist even after the malicious participant exits the federation, all while maintaining nominal clean-task utility.

The evolution of literature within these subtypes reflects a continuous arms race against specific defence paradigms. In global gradient poisoning, the methodology shifts from brute-force directional consistency~\cite{xie2025model} to sophisticated statistical mimicry~\cite{yang2025enhanced} designed to defeat sign-based monitoring. At the coordinate level, the focus evolves from optimizing sparse fragments for efficiency~\cite{ren2024fmpa} to geometrically bypassing trusted anchors using dual-objective optimization~\cite{Kasyap2024Sine}. Finally, the backdoor domain bifurcates based on topology: in VFL settings, research targets the "label invisibility" constraint through gradient-based inference and semantic alignment~\cite{Chen2024Universal,Yang2025Stealthy}, whereas in HFL, the primary innovation lies in circumventing robust aggregation via density-adaptive triggers and projected gradient updates~\cite{wang2025improved}. This progression underscores that financial FL security cannot rely on static, single-metric defenses, but requires adaptive auditing capable of detecting both statistical anomalies and functional rewrites.

\begin{table*}[t]
\small
\centering
\caption{Comparison of model poisoning subtypes in the training phase, highlighting algorithmic strategy, computational profile, evasion mechanism, persistence characteristics, target financial tasks, and downstream operational consequences in distributed financial learning.}
\label{tab:model-poisoning-summary}
\renewcommand{\arraystretch}{1.3}
\setlength{\tabcolsep}{6pt}
\begin{tabularx}{\textwidth}{l
                            >{\raggedright\arraybackslash}X
                            >{\raggedright\arraybackslash}X
                            >{\raggedright\arraybackslash}X}
\toprule
\textbf{Dimension} &
\textbf{b1 Global Gradient Poisoning} &
\textbf{b2 Fragment/Coordinate-Level Poisoning} &
\textbf{b3 Backdoor/Trigger-Based Poisoning} \\
\midrule
\multicolumn{4}{c}{\textit{Algorithmic \& Methodological Contrast}} \\
\midrule
\textbf{Algorithmic Strategy} &
Cross-round directional consistency~\cite{xie2025model} vs. statistical sign-distribution mimicry~\cite{yang2025enhanced}. &
Sparse high-leverage coordinate perturbation~\cite{ren2024fmpa} vs. geometric anchor-similarity bypass~\cite{Kasyap2024Sine}. &
Clean-label gradient inference (VFL)~\cite{Chen2024Universal,Yang2025Stealthy} vs. density-adaptive sub-triggers (HFL)~\cite{wang2025improved}. \\

\textbf{Complexity Profile} &
Linear $O(d)$ per round for directional scaling; sorting overhead for sign matching. &
Scales with fragment size $k \ll d$ (local crafting); efficient for high-dimensional models. &
Requires auxiliary data training (VFL) or projection steps (HFL); inference overhead is negligible. \\

\textbf{Evasion Mechanism} &
Dynamic magnitude scheduling to pass outlier checks~\cite{xie2025model}; matching benign sign ratios~\cite{yang2025enhanced}. &
Concentrating damage on specific layers while maintaining global vector norms~\cite{ren2024fmpa}. &
Semantic alignment to target clusters~\cite{Yang2025Stealthy}; projected gradients within benign boundaries~\cite{wang2025improved}. \\

\textbf{Stealth \& Persistence} &
Medium; relies on slow accumulation of small deviations across multiple rounds. &
High; specifically targets blind spots in single-metric defenses (e.g., cosine only). &
Very High; functional rewrite persists even if the attacker leaves; triggers are hidden in normal inputs. \\

\midrule
\multicolumn{4}{c}{\textit{Financial Deployment \& Systemic Impact}} \\
\midrule
\textbf{Target Financial Task} &
High-frequency trading and real-time fraud detection aggregating rapid updates~\cite{yang2025enhanced}. &
Consortium credit profiling and AML models with non-IID data distributions~\cite{Kasyap2024Sine}. &
Cross-institution VFL credit scoring~\cite{Chen2024Universal} and multi-bank HFL risk screening~\cite{wang2025improved}. \\

\textbf{Downstream Consequence} &
General convergence failure or random-guess degradation in joint models. &
Persistent misclassification on specific risk regions or user segments; detection evasion. &
Targeted approval of high-risk loans; "safe passage" for laundering via hidden triggers. \\

\textbf{Systemic Risk Priority} &
Medium &
High &
Very High \\
\bottomrule
\end{tabularx}
\end{table*}

\section{Model Deployment and Inference Phase}
\label{sec:model-deploy}

Financial AI models operate on highly dynamic and non-stationary time series as well as structured transaction data, which together instantiate multiple attack dimensions defined in the taxonomy outlined above. Rather than exhibiting isolated vulnerabilities, financial systems expose distinct attack surfaces that align with different axes of adversarial perturbation design, including update strategy, attacker knowledge, constraint structure, and attack objective.

In trading systems, commonly deployed architectures such as LSTM, CNN, and RNN—designed to capture temporal dependencies—naturally amplify gradient-based perturbations, making both single-step and multi-step attacks particularly effective. Subtle, norm-bounded manipulations to order book entries can propagate through recursive temporal updates, distorting perceived supply–demand dynamics and triggering cascades of erroneous trading decisions. Such behaviours exemplify how perturbation construction mechanisms interact with temporal model structure\cite{zhang2019extending,goldblum2021adversarial}.

In credit scoring and fraud detection, adversarial behaviour more frequently manifests as feature-level perturbations under explicit constraint models, where attackers adjust transaction amounts, merchant codes, or timestamps to evade detection while preserving semantic plausibility. These attacks are often realised through optimisation-based perturbations, which explicitly trade off evasion effectiveness against detectability, or through transfer-based strategies that exploit surrogate models to bypass black-box decision systems commonly deployed in practice~\cite{kazoom2025boundary,cartella2021adversarial}.

These effects are further magnified in fully automated and high-frequency trading environments, where millisecond-level execution removes human-in-the-loop safeguards. In such settings, learning-based attacks, including transfer-based and reinforcement-learning-driven perturbations, can accumulate effects across decision rounds and propagate across interacting models\cite{chen2021adversarial}. Flash Crash–like events~\cite{kirilenko2017flash} illustrate how adversarial perturbations, when aligned with temporal accumulation and automation, may escalate from local manipulation to systemic instability.

Collectively, these observations demonstrate that adversarial risks in financial AI systems cannot be adequately understood through isolated attack examples alone. Instead, they motivate a taxonomy-driven perspective that disentangles how perturbations are constructed, what knowledge attackers assume, which constraints govern feasible manipulations, and what objectives are optimised. Such a structured view is essential for systematically analysing, comparing, and mitigating adversarial threats in real-world financial AI deployments.

\subsection{Adversarial Attacks}
\label{sec:adversial-attacks}

Adversarial attacks refer to carefully crafted perturbations to model inputs that induce misclassification, erroneous predictions, or unstable behaviours in AI-based systems. Unlike natural noise, adversarial perturbations are designed with a precise intent to exploit vulnerabilities in model architectures, training procedures, and decision boundaries. In the finance domain, the motivation for studying adversarial attacks lies in the dual need to Both assess the robustness and reliability of time-critical financial models under worst-case conditions, and to anticipate and prevent malicious manipulation that may distort markets, enable unfair profit, or compromise systemic stability.

\subsubsection{Input Perturbation Attacks}
\label{sub:input_perturbation_attacks}

Fursov et al.\ (2021) investigate the fragility of 1D-CNN-based financial time-series classifiers under highly localised adversarial perturbations~\cite{fursov2021adversarial}. Motivated by the observation that financial signals exhibit strong temporal locality, the authors propose the \emph{Single Value Attack} (SVA), which adapts the Fast Gradient Sign Method (FGSM) to perturb only the most influential timestep within a sliding window rather than distributing noise across all entries.

Recognising that full-window perturbations may be detectable under volatility constraints, SVA computes gradients with respect to each timestep in a window $x \in \mathbb{R}^n$ and identifies the index $i^*$ maximising the absolute gradient magnitude. A bounded perturbation is then applied exclusively to $x_{i^*}$, ensuring that the modification remains within empirically observed price-delta ranges.

The method is evaluated on Google stock data from 2006 to 2018 using 30-day sliding windows to predict 14-day future movement. Results demonstrate that perturbing a single entry can flip directional predictions while preserving statistical plausibility. Computationally, the attack requires one backward pass and $n$ forward evaluations per window, yielding linear complexity in window size.

While SVA reveals concentrated temporal sensitivity in CNN-based forecasting models, it assumes white-box gradient access and does not evaluate cross-model transferability, leaving open questions regarding robustness under black-box constraints.

Pandey et al.\ (2023) address the limitations of traditional $\ell_p$-norm adversarial constraints in heterogeneous financial tabular data~\cite{pandey2023improving}. Observing that feature scales in fraud detection vary significantly (e.g., transaction amount versus binary indicators), they argue that $\ell_0$-oriented perturbation strategies better reflect practical evasion settings.

To identify scale-invariant vulnerable features, they propose a GAN-based \emph{Feature Selector Network} (FSN) that learns a binary mask $m \in \{0,1\}^D$ indicating a minimal subset of features sufficient for successful evasion. The generator learns to concentrate perturbations on mid-rank features—those influential enough to affect classification but less scrutinised than high-importance features.

Experiments on the IEEE-CIS and European Credit Card Fraud datasets demonstrate that restricting black-box attacks (Boundary, HopSkipJump) to FSN-selected features reduces query complexity by approximately 75\% and attack generation time by 50\%, while improving imperceptibility measured via nearest-neighbour distance metrics. Adversarial training using FSN-generated samples reduces attack success rates by roughly 40\%.

Although FSN improves efficiency in classification-based fraud detection, it does not extend to multi-objective or regression-based financial settings where feature interactions are structurally constrained.

Raff et al.\ (2025) extend input-based adversarial perturbations to structured financial reporting systems, where predictive tasks are regression-based and constrained by accounting identities~\cite{raff2025adversarial}. Unlike prior classification-focused evasion attacks, their work formalises manipulation of financial disclosures as a constrained multi-objective optimisation problem.

Motivated by the observation that distressed firms may simultaneously seek to inflate Earnings Per Share (EPS) while reducing fraud-detection indicators such as the Beneish M-score, the authors introduce the Maximum Violated Multi-Objective (MVMO) attack. The method employs a softmax-gated loss function that dynamically emphasises the objective exhibiting the largest deviation from its target, thereby balancing competing regression goals during optimisation.

To address the structural dependencies inherent in financial ratios, the attack is implemented using Projected Gradient Descent within an automatic differentiation framework. Because ratio-based accounting metrics introduce denominator coupling across reporting periods, the resulting objective function is non-convex. Furthermore, perturbations must preserve balance-sheet consistency and inter-period accounting relationships, significantly constraining the feasible action space.

Empirical evaluation on simulated financial reporting scenarios indicates that the method achieves approximately 50\% success in jointly satisfying earnings inflation and fraud-score reduction objectives under bounded perturbation budgets. The study demonstrates that multi-task regression models used in financial oversight may exhibit vulnerability to structurally consistent adversarial manipulation.

However, the attack assumes gradient access to the reporting model and requires domain-informed feasibility constraints to ensure accounting plausibility. It does not evaluate cross-model transferability or robustness under independent regulatory auditing pipelines, leaving broader systemic implications as an open research question.

Collectively, these studies demonstrate that input-based perturbations in financial AI evolve from highly localised gradient manipulation in time-series models, to scale-invariant feature subset optimisation in tabular classification, and finally to structurally constrained multi-objective optimisation in reporting regression models. While each operates at inference time, the complexity of feasible perturbations increases with domain structure, highlighting the necessity of domain-aware robustness evaluation beyond conventional $\ell_p$-bounded adversarial testing.

\subsubsection{Single-step adversarial example generation}
\label{par:single_step_perturbations}

Single-step perturbations craft adversarial inputs by applying a single gradient-aligned update to maximize prediction loss under a bounded perturbation budget: $\mathbf{x}_{\text{adv}}=\mathbf{x}+\epsilon\cdot\mathrm{sign}\!\left(\nabla_{\mathbf{x}}L(\mathbf{x},y)\right)$. Unlike poisoning attacks that alter training data, these attacks operate at inference time and require only one backward pass, yielding linear time and space complexity $O(N)$.

In financial systems, such perturbations can be calibrated to resemble natural volatility or benign transaction noise, making them particularly stealthy in deployment environments.

The conceptual foundation of single-step perturbations originates from the linearity hypothesis proposed by Goodfellow et al.~\cite{goodfellow2014explaining}. Contrary to earlier conjectures attributing adversarial vulnerability to extreme non-linearity or overfitting, the authors argue that the phenomenon arises from the locally linear behaviour of neural networks in high-dimensional spaces. Modern architectures such as ReLU networks and LSTMs are intentionally designed to exhibit near-linear responses to facilitate gradient-based optimisation. However, this local linearity implies that small but coordinated perturbations can accumulate additively across input dimensions, leading to disproportionately large shifts in activations.

Formally, the Fast Gradient Sign Method (FGSM) constructs adversarial perturbations as $\eta = \epsilon\,\mathrm{sign}(\nabla_x J(\theta,x,y))$. Under a linear approximation, the perturbed activation satisfies $w^\top \tilde{x} = w^\top x + w^\top \eta$, where for an $n$-dimensional weight vector with average magnitude $m$, the activation shift scales as $\epsilon m n$. Thus, even infinitesimal perturbations $\epsilon$ can produce substantial output changes when aligned with the gradient direction in high-dimensional input spaces.

Empirical evaluations demonstrate that such single-step perturbations induce high-confidence misclassification across standard benchmarks, while requiring only a single backpropagation pass. This computational efficiency makes both large-scale attack deployment and adversarial training feasible. Notably, incorporating FGSM-generated samples into the training objective significantly improves robustness without prohibitive computational overhead, establishing single-step perturbations as both an attack mechanism and a practical robustness evaluation tool.

While Goodfellow's work demonstrated the effectiveness of single-step perturbations on the MNIST dataset, Gallagher et al.\ (2022) extend this framework to financial time-series forecasting using LSTM and CNN models trained on stock data spanning 2006–2018~\cite{gallagher2022investigating}. In their setup, perturbations are applied to 30-day sliding windows with magnitudes $\epsilon$ calibrated to typical daily market volatility to maintain realism. As the perturbation scale increases, the models exhibit sharply rising binary cross-entropy loss accompanied by significant drops in accuracy, precision, recall, and F1 scores. Simulated backtesting further reveals that these adversarially perturbed forecasts cause monotonic declines in cumulative profits, reaching losses of approximately \$200k under moderate noise levels. This outcome illustrates that even small, distributed perturbations—applied across multiple timesteps—can exploit the smooth structure of financial signals to systematically distort directional predictions and realised P\&L.


While the above scenario focuses on stock time-series forecasting, adversarial vulnerabilities also arise in transaction-level financial records. Fursov et al. (2021) investigate transaction-level financial models under both white-box and black-box threat settings~\cite{fursov2021adversarial}. Rather than perturbing continuous-valued inputs, their approach operates on tokenised transaction sequences, where embedding-level gradient updates are computed and subsequently projected back to valid discrete tokens.

The authors propose these adversarial attack vectors: Simpling Fool(SF), Concat-FGSM, which appends adversarial tokens to the sequence, and LM-FGSM, which leverages a language model to guide token substitution. Experimental results demonstrate strong vulnerability: modifying only one or two transaction records achieves attack success rates exceeding 0.8 in black-box settings, while white-box attacks reduce model accuracy to near zero.

These findings suggest that transaction-level models are highly sensitive to sparse structural perturbations, even when the number of altered records is minimal. Importantly, all proposed methods maintain $O(N)$ time and space complexity, preserving computational scalability for large transaction sequences.

In contrast to time-series attacks that rely on distributed continuous perturbations across multiple timestamps, transaction-level attacks can alter model predictions by modifying only a small number of discrete transaction records.

Both time-series and transaction-level studies operate at the feature level, yet they differ in input carriers and optimisation schemes. Gallagher et al.\ leverage continuous numerical features in sliding windows and apply gradient-aligned perturbations directly, producing distributed shifts across many timesteps. This approach scales linearly with window size $O(N)$ and is particularly effective at distorting temporally recursive indicators such as price trends and realized P\&L. In contrast, Fursov et al.\ manipulate discrete tokenised transactions using embedding-level gradients projected back to valid tokens, relying on sparse structural edits. Although only a few tokens are altered, attacks maintain $O(N)$ complexity and can achieve high success rates even in black-box settings, making them suitable for network-like or combinatorial tasks such as fraud detection, account risk evaluation, or merchant clustering.

The two paradigms differ in operational impact: continuous perturbations degrade forecasts incrementally but systematically, while sparse discrete edits induce abrupt decision flips. Neither approach yet incorporates effective detection or defensive mechanisms, and open questions remain regarding cross-model generalisation, dynamic adaptation, and robustness to real-time financial system evolution. This contrast clarifies efficiency bottlenecks and task–attack fit for feature-level single-step perturbations, providing a framework to prioritise defence strategies according to input modality and operational context.

\subsubsection{Optimisation-based adversarial example generation}


Optimisation-based adversarial attacks formulate adversarial example generation as a constrained optimisation problem that explicitly seeks a minimal perturbation that causes a model misclassification. The mathematical statement of this objective is $\min_{\delta}\; \| \delta \|_p \quad\text{subject to}\quad C(x+\delta)=t,\qquad x+\delta \in \mathcal{X}$. where $x$ denotes a valid input, $\delta$ denotes the perturbation, $C$ denotes the classifier's decision rule, $t$ a specified target label, and $\mathcal{X}$ the domain of valid inputs. The choice of the $L_p$ norm in the objective encodes a precise notion of perturbation magnitude; common choices in the literature include the $L_0$, $L_2$ and $L_\infty$ norms. Under this formulation, adversarial robustness can be interpreted as the minimal distance from an input to the classifier's decision boundary measured in the selected norm.

Early optimisation-based adversarial attacks formalised adversarial examples construction as a box-constrained minimal-perturbation problem. Szegedy et al.\ first demonstrated that solving a penalised $L_2$ optimisation can produce imperceptible yet targeted perturbations~\cite{szegedy2013intriguing}. Their approach approximates the original constrained problem using a penalty function method, and applies the L-BFGS-B quasi-Newton solver to optimise $\min_{x' \in [0,1]^n} \; c \cdot \|x - x'\|_2^2 \;+\; \mathrm{loss}_{F,l}(x')$, where the penalty parameter $c$ is selected via line search to balance misclassification success with perturbation magnitude. This formulation provided the first constructive evidence that deep neural networks exhibit highly non-smooth and discontinuous input--output mappings, enabling adversarial examples that transfer across architectures and independently trained models~\cite{szegedy2013intriguing}.

Madry et al.\ (2019)~\cite{madry2017towards} proposed the \emph{Projected Gradient Descent (PGD)} attack, it iteratively updates perturbations along the gradient and projects them back into the admissible set (typically an $\ell_\infty$ ball of radius $\epsilon$). The method reframes adversarial robustness as a robust optimisation problem, where the attacker maximises loss within the perturbation budget, and the defender minimises worst–case loss. Experiments on MNIST and CIFAR-10 showed that PGD-trained models retain high accuracy under strong first-order white-box attacks (e.g., $>89\%$ on MNIST, $>45\%$ on CIFAR-10), clearly outperforming the standard. Although PGD has a slight decrease in accuracy compared to FGSM on clean data, adversarial training with PGD significantly improves robustness against both white-box and transfer-based black-box attacks. These results support PGD as a strong first-order adversary and a standard baseline for adversarial training.

PGD lays the theoretical foundation for optimisation-based adversarial example generation: on one hand, its formulation of $\epsilon$-bounded constraints and projection at each step improves attack efficiency; on the other hand, the optimisation framework naturally extends to more complex domains such as adversarial manipulation in high-frequency trading systems.

Battista et al.\ (2021)~\cite{goldblum2021adversarial} adapt PGD-based gradient ascent to order-book forecasting in high-frequency trading (HFT), where inputs are order-book snapshots and streaming features aggregated over 60-second windows ($N_{\text{ob}}$ events per slice). Perturbations must respect market microstructure constraints including bounded capital budget $c$, integer-valued order sizes, and nonnegative volume additions without directly altering prices; at each step, updates follow the loss gradient with randomised step size $\alpha_k \in (0, \alpha_0]$ and are projected back to the feasible set via budget accounting and integer rounding. Per-attack cost is $O(k\,N_{\text{ob}})$ time and $O(N_{\text{ob}})$ space, with $k$ gradient-ascent steps. Evaluation on MLP, LSTM, and linear regression forecasters shows all models suffer reduced robustness: MLPs are less robust than linear models despite higher clean-data accuracy, while LSTMs exhibit gradient masking that resists direct attacks but leaves them highly vulnerable to transfer attacks. Notably, equities with high price per share and low trading volume prove harder to attack, as integer rounding yields coarse admissible perturbations that reduce gradient-based efficacy.

However, compared to traditional LP norm-based optimisation methods, Reinforcement learning is gradually becoming a more effective optimisation method. Lunghi et al. proposed FRAUD-RLA~\cite{lunghi2025fraud}, which introduces a novel RL framework designed to bypass sophisticated fraud classifiers. The paper bridges the gap between general adversarial theory and financial applications by recognising that fraudsters operate with limited information and must maximise their success before a Fraud Detection Engine (FDE) adapts or the card is neutralised. Unlike previous attacks that require deep access to a victim’s device, FRAUD-RLA optimises the exploration-exploitation tradeoff to identify successful fraud patterns rapidly and with minimal prior knowledge. 

\subsubsection{Universal Transfer-Based perturbations}

Transfer-based attacks exploit the empirical phenomenon of \emph{adversarial transferability}, namely that adversarial examples crafted on one model often remain effective against other models trained on the same data distribution~\cite{szegedy2013intriguing, papernot2016transferability}. Formally, let $f_s$ denote a surrogate model and $f_t$ the unknown target model. An attacker optimises a perturbation $\delta$ such that $\arg\max_{\|\delta\|\le\epsilon} \mathcal{L}(f_s(x+\delta), y)$ and deploys $\delta$ against $f_t$, relying on shared gradient alignment or similar decision boundaries.

Universal Adversarial Perturbations (UAPs) strengthen this threat by identifying a single vector $\delta_u$ satisfying $\Pr_{x \sim \mathcal{D}}\left(f(x+\delta_u) \neq f(x)\right) \geq \eta$, for a large fraction $\eta$ of the data distribution~\cite{moosavi2017universal}. Unlike per-instance attacks (e.g., FGSM~\cite{goodfellow2014explaining}, C\&W~\cite{carlini2017towards}), UAPs enable distribution-level manipulation, substantially lowering attack cost and increasing scalability.

In financial systems, where model architectures are typically proprietary, transfer-based methods operationalise realistic black-box threat models~\cite{papernot2016transferability}. This is particularly relevant in Model Risk Management (MRM), where reliance on architectural secrecy is common but theoretically insufficient.

The experimental framework utilised by Piazza et al. evaluates a Basic DQN and a TensorTrade DQN mimicking real-world BTC/USD trading~\cite{piazza2021adversarial}. Applying Fast Gradient Sign Method (FGSM) and Carlini \& Wagner (C\&W) attacks ~\cite{carlini2017towards} revealed that these agents are acutely sensitive to perturbations in the most recent window history tuple of their features. A critical finding for risk managers is that the resulting net-worth impact is often not reflected in the total reward metric, potentially masking the severity of an attack from human operators. Furthermore, the research demonstrates high transferability to target agents; attacks crafted on imitated agents (even those trained on "imperfect" or non-continuous demonstrations) successfully transferred to the proprietary target, dropping performance significantly. This highlights that test-time efficiency is achievable through imitation learning, allowing adversaries to bypass the need for direct model access.

The above evidence in DRL-based trading systems illustrates transferability in sequential decision settings, where perturbations crafted on surrogate policies generalise across agents trained under similar Markov Decision Processes. Importantly, this vulnerability is not unique to reinforcement learning. Rather, it reflects a broader geometric property of models trained on shared data distributions: decision boundaries across heterogeneous architectures often exhibit substantial alignment.

\begin{table}[htbp]
\small
\centering
\caption{Comparison of adversarial attack categories in the model deployment and inference phase, summarizing their methodological structure, access assumptions, perturbation patterns, and downstream operational impact in financial AI systems.}
\label{tab:adv-attacks-summary}
\renewcommand{\arraystretch}{1.3}
\setlength{\tabcolsep}{5pt}
\begin{tabularx}{\textwidth}{l
                            >{\raggedright\arraybackslash}X
                            >{\raggedright\arraybackslash}X
                            >{\raggedright\arraybackslash}X
                            >{\raggedright\arraybackslash}X}
\toprule
\textbf{Dimension} &
\textbf{(a1) Input Perturbation} &
\textbf{(a2) Single-Step Generation} &
\textbf{(a3) Optimisation-Based Generation} &
\textbf{(a4) Universal / Transfer-Based} \\
\midrule
\multicolumn{5}{c}{\textit{Methodological Structure}} \\
\midrule

\textbf{Core Objective} &
Task-specific inference-time manipulation under domain constraints~\cite{fursov2021adversarial,pandey2023improving,raff2025adversarial}. &
One-step gradient-aligned loss maximisation (FGSM)~\cite{goodfellow2014explaining}. &
Explicit constrained optimisation for minimal or feasible perturbations~\cite{szegedy2013intriguing,madry2017towards}. &
Cross-model or distribution-level misclassification without target access~\cite{papernot2016transferability,moosavi2017universal}. \\

\textbf{Representative Works} &
SVA~\cite{fursov2021adversarial}; FSN~\cite{pandey2023improving}; MVMO~\cite{raff2025adversarial}. &
FGSM~\cite{goodfellow2014explaining}; financial time-series applications~\cite{gallagher2022investigating}. &
L-BFGS~\cite{szegedy2013intriguing}; PGD~\cite{madry2017towards}; HFT-constrained PGD~\cite{goldblum2021adversarial}. &
Transferability~\cite{papernot2016transferability}; UAP~\cite{moosavi2017universal}; DRL transfer~\cite{piazza2021adversarial}; CCFD transfer~\cite{fok2025foe}. \\

\textbf{Optimisation Scheme} &
Gradient-based but domain-aware (feature masking, structural feasibility). &
Single backward pass; linear complexity. &
Iterative projected gradient or solver-based optimisation. &
Surrogate training + deployment on unseen target. \\

\textbf{Access Assumption} &
Primarily white-box. &
White-box gradient access. &
White-box first-order or solver-level access. &
Black-box feasible via transferability or imitation learning. \\

\textbf{Perturbation Structure} &
Localised timestep edits, sparse feature subsets, or structurally constrained accounting-consistent modifications~\cite{fursov2021adversarial,pandey2023improving,raff2025adversarial}. &
Distributed small-magnitude continuous perturbations aligned with input gradients~\cite{goodfellow2014explaining,gallagher2022investigating}. &
Norm-minimal or boundary-aligned perturbations obtained via iterative projection or constrained optimisation~\cite{szegedy2013intriguing,madry2017towards,goldblum2021adversarial}. &
Cross-model transferable perturbations or single universal vectors effective across input distributions~\cite{papernot2016transferability,moosavi2017universal,piazza2021adversarial,fok2025foe}. \\

\midrule
\multicolumn{5}{c}{\textit{Financial Operational Impact}} \\
\midrule

\textbf{Primary Financial Surface} &
Time-series forecasting; fraud detection; reporting regression. &
Time-series and transaction-level classification. &
High-frequency trading; structured financial models. &
DRL trading agents; fraud ensembles; proprietary pipelines. \\

\textbf{Integrity Impact Pattern} &
Prediction flips under plausibility constraints. &
Systematic metric degradation with low computational cost. &
Worst-case loss amplification within feasibility region. &
Cross-architecture compromise; distribution-wide impact. \\

\textbf{Scalability} &
Moderate; depends on structural constraints. &
High; $O(N)$ per sample. &
Lower; $O(kN)$ or solver-dependent. &
High after perturbation construction (constant marginal cost for UAP). \\

\textbf{Threat Scope} &
Model-specific exploitation. &
Local linear vulnerability. &
Decision-boundary proximity exploitation. &
Distribution-level and cross-model vulnerability propagation. \\

\bottomrule
\end{tabularx}
\end{table}

This cross-model alignment becomes even more explicit in tabular financial domains such as credit card fraud detection (CCFD), where models operate on fixed feature vectors rather than temporal state windows. In these settings, transferability manifests as \emph{cross-architecture} vulnerability, whereby perturbations generated on gradient-based models degrade the performance of structurally distinct ensemble methods.


Foka et al.\ investigate cross-architecture transferability using FGSM to generate adversarial samples on a Logistic Regression (LR) surrogate attacking a Random Forest (RF) target~\cite{fok2025foe}. On the Kaggle CCFD dataset of European cardholders ($0.17\%$ fraud rate), with SMOTE for class imbalance and the Adversarial Robustness Toolbox (ART), the white-box attack on LR causes recall to drop from $92\%$ to $56\%$ and precision from $0.17$ to $0.11$. These perturbations transfer with $94\%$ efficacy to the RF model, suggesting shared decision boundaries learned from the same data distribution and undermining the defensive strategy of heterogeneous model ensembles. Complexity analysis reveals that PCA-transformed features (V1--V28) are highly sensitive to consistent perturbations of approximately $\pm 2.2$, beyond which a saturation effect yields no further degradation, implying a ceiling for FGSM-based effectiveness in the tabular domain as samples move beyond the effective decision boundary into unrealistic distributions.

The financial sector faces an adversarial landscape characterised by a diversity of attack surfaces: high-dimensional DRL state spaces, discrete tabular fraud features, and raw time-series signals. Framing these threats through the CIA triad reveals that while imitation attacks target policy confidentiality, transferable and universal perturbations primarily undermine the integrity of automated decision-making.

\subsubsection{Comparative Analysis of Adversarial Attack Subtypes}

Table~\ref{tab:adv-attacks-summary} systematises adversarial attacks in financial AI along algorithmic mechanisms, access assumptions, complexity profile, and systemic risk gradient. 
While single-step perturbations exploit local linearity with minimal computational overhead, iterative and optimisation-based attacks strengthen boundary alignment through multi-step or solver-based procedures, increasing attack reliability at a higher cost. 
Universal and transfer-based attacks extend this threat surface by decoupling attack generation from target-model access, enabling scalable disruption across black-box financial systems.

Across all subtypes, the dominant security concern lies in integrity compromise rather than confidentiality leakage. 
In time-series forecasting and trading agents, adversarial perturbations directly translate into realised capital loss. 
In fraud detection and credit scoring, they induce systematic misclassification that may enable illicit financial flows or credit misallocation. 
Notably, universal and cross-architecture transfer attacks represent the highest systemic risk, as they undermine the defensive assumption that model heterogeneity or parameter secrecy provides protection.

The progression from single-step to optimisation-based and ultimately transfer-driven attacks reflects an arms race between efficiency and robustness in financial AI. 
Early gradient-aligned methods revealed linear vulnerabilities; robust optimisation reframed adversarial robustness as a min–max problem; and transfer-based approaches demonstrated that black-box opacity does not guarantee security. 
This trajectory suggests that financial model risk management must move beyond isolated robustness metrics toward adaptive, distribution-aware auditing frameworks capable of detecting both local perturbations and cross-model vulnerability propagation.

\section{Model Monitoring and Maintenance Phase}
\label{sec:model-monitoring}

The model monitoring and maintenance phase represents the operational steady-state where AI systems actively interact with users, execute automated workflows, and adapt to emerging data streams. Unlike the training phase, which focuses on optimization, or the inference phase, which focuses on decision boundaries, this stage is characterized by continuous, open-ended interaction with external agents and environments. In modern FinTech architectures, this phase increasingly relies on Large Language Models (LLMs) to function as autonomous agents and on biometric systems to verify identity. However, this operational autonomy introduces a critical vulnerability: the system's reliance on the semantic integrity of its inputs. Because these models are designed to follow natural language instructions and accept high-fidelity sensory data as ground truth, they effectively lack a distinction between malicious manipulation and legitimate interaction. Adversaries exploit this "semantic trust" not by corrupting mathematical weights, but by hijacking the high-level intent of the system or by fabricating the evidentiary basis of identity. This creates two distinct threat vectors: manipulating the communicative logic via prompt injection\cite{bayhan2025prompt}, or synthesising fraudulent biological credentials via deepfakes\cite{ahmad2024survey}.

\subsection{Prompt Injection}

Prompt injection constitutes a security failure in AI-mediated workflows where adversaries manipulate input streams to override the system's original operating instructions. Fundamentally, this attack exploits the \emph{control-data confusion} inherent in LLM architectures, where user inputs (data) and system instructions (control) share the same communication channel\cite{perez2022ignore}. By embedding adversarial directives—such as ``ignore previous rules'' or ``act as an unrestricted administrator''—into the input context, attackers can force the model to prioritize malicious commands over its safety alignment.

In financial contexts, this mechanism allows adversaries to hijack the execution logic of AI agents that possess API access or decision-making authority. For instance, in automated customer service, an injected prompt can coerce a virtual assistant into revealing sensitive internal protocols or other users' transaction histories (Data Leakage). In algorithmic trading or portfolio management, attackers may inject misleading context into news feeds or analyst reports processed by the AI, inducing the model to trigger erroneous buy/sell orders (Business Logic Interference)\cite{zhan2025adaptive}. Furthermore, internal code-generation assistants can be manipulated to output insecure SQL queries or bypass compliance checks\cite{pearce2025asleep}. Unlike traditional software exploits that require binary vulnerability, prompt injection operates entirely at the semantic level, making it particularly difficult to filter using conventional firewalls or keyword matching, as the malicious payload is syntactically indistinguishable from legitimate natural language.

\subsubsection{Character/Symbol-Level Injection}

Character/symbol-level injection makes \emph{tiny edits to an otherwise legitimate prompt string} to bypass guardrails and induce sensitive disclosure or over-privileged actions. This subsection contrasts two families—zero-width insertion and homoglyph/digit mapping—in white/black-box settings, analyzing their mechanisms and deployment implications.

Sang et al.\ (2024) evaluate safety under injection where inputs may only receive character-level perturbations while preserving task semantics using Microsoft PromptBench~\cite{sang2024evaluating}. The study pressure-tests guardrails with minimal string rewrites such as DeepWordBug\cite{gao2018black} and TextBugger\cite{li2018textbugger}. Per string, construction is a linear scan $O(n)$; composing $k$ transforms is $O(k \cdot n)$. Results show Claude outperforming Mistral Large across all percentages (e.g., accuracy 92.5\% vs 89.8\%). While this study effectively quantifies the model's resilience against standard textual perturbations (e.g., typos and swaps), the character-level attack surface extends beyond visible noise to more sophisticated encoding manipulations targeting the detection layers.

Expanding this category to encoding-level obfuscation, Hackett et al.\ (2025) investigate how non-standard character representations can bypass guardrails entirely while remaining intelligible to the LLM~\cite{hackett2025bypassing}. This work systematically rewrites characters/symbols to evade multiple prompt-injection and jailbreak detectors. The mechanism exploits segmentation/feature-extraction brittleness to encoding/rendering changes while LLMs still parse payloads; techniques include ``emoji smuggling'' and ``Unicode tag smuggling.'' A single technique scans the string in $O(n)$. Reported bypass rates are high: emoji smuggling achieves 100\% on two datasets. Because these transforms are extremely cheap, channel-agnostic, and often slip through preprocessing, they directly transfer to finance, such as eliciting sensitive context in advisor chats or wrapping triggers with Unicode in KYC flows to sidestep guards.

The synthesis of this subtype reveals a vulnerability in the \emph{encoding/rendering to tokenization mismatch}. Attacks utilize minute character/symbol rewrites (zero-width, homoglyphs, bidi control) that keep semantics readable while perturbing tokenization, regexes, and normalization. Entry typically needs no external data or tools and works in black-box mode. Financially, cross-system text paths—advisor chats, ticketing, compliance summaries—are prioritized where inconsistent normalization amplifies penetration.

\subsubsection{Word/Phrase-Level Injection}

Word/phrase-level injection exploits the \emph{semantic and visual processing gaps} of LLMs by manipulating discrete tokens—keywords, phrases, or formatting codes—to trigger unintended behaviors while maintaining surface-level legitimacy. Unlike prompt-level overwrites, these attacks are surgical: they target specific recognition modules or decision logic.

Liu et al.\ (2025) introduce \emph{ChameleonAttack}, a semantics-preserving injection framework designed for event-driven stock prediction~\cite{liu2025semantics}. Addressing the challenge that discrete token optimization often produces gibberish, the authors propose a two-stage mechanism: searching for a malicious vector in the continuous latent space that maximizes prediction error, then using a ``Semantic-Translation Module'' to map this signal back into natural financial news. However, while achieving high linguistic quality, this approach relies on iterative gradient computation and auxiliary translation models, imposing significant computational overhead ($O(N)$ inference cost) that limits applicability in high-frequency trading.

Addressing the computational bottlenecks of gradient-based generation, Rizvani et al.\ (2026) pivot towards an operationally lightweight paradigm focused on visual deception with \emph{Adversarial News}~\cite{rizvani2026adversarial}. This method exploits the ``parsing gap'' using low-cost primitives like Unicode Homoglyphs and Hidden-Text Injection. By altering a few bytes in a headline, an attacker can misroute a positive news signal to be ignored by the ATS. The complexity is negligible ($O(1)$ string replacement), making it scalable. Nevertheless, because these attacks depend on static patterns, they remain brittle against rule-based sanitization filters.

To overcome the limitations of static heuristics and bypass simple sanitization, Shi et al.\ (2024) reintroduce optimization-based principles but apply them to the decision-making layer with \emph{JudgeDeceiver}~\cite{shi2024optimization}. This method formulates injection as a constrained optimization problem: finding a distinct token sequence (suffix) that forces a ``LLM-as-a-Judge'' to rank a target response as the best option. The algorithm uses gradient-based search to minimize a composite loss function balancing target alignment with stealth. In finance, where LLMs increasingly act as automated compliance officers, this method exposes the ability to mathematically rig the jury.

The comparison of these studies illustrates a spectrum from computationally intensive semantic optimization to lightweight structural obfuscation. Liu et al.\ and Shi et al.\ prioritize attack capability through gradient-based search ($O(N)$), whereas Rizvani et al.\ prioritize operational efficiency ($O(1)$) via static character-level disparities. This dichotomy maps to financial vulnerabilities: high-overhead semantic approaches threaten asynchronous, high-value workflows (compliance auditing), while low-overhead structural attacks imperil synchronous, latency-sensitive systems (HFT news feeds).

\subsubsection{Sentence/Instruction-Level Injection}

Sentence/instruction-level injection directly inserts, rewrites, or splices \emph{complete commands} so the model treats them as high-priority instructions (e.g., ``ignore prior constraints''), often nudging the task over multiple dialogue turns.

Jiang et al.\ (2023) introduce \emph{Prompt Packer}, a framework executing Compositional Instructions Attacks (CIA) by encapsulating harmful intents within benign-looking task instructions~\cite{jiang2023prompt}. The strategy leverages composition and encapsulation to hide the harmful goal inside a harmless template (e.g., dialogue or writing tasks) to bypass refusal. Complexity grows linearly with generation steps and instruction length, $O(T \times n)$. Results show high success (ASR 95\%+ on benchmarks). However, it remains confined to static textual logic, which invites the exploration of more dynamic, cross-modal injection vectors.

Expanding the concept of encapsulation to the temporal and multimodal domain, Chiu et al.\ (2025) address voice interaction capabilities with \emph{Flanking Attack}~\cite{chiu2025say}. In the context of financial voice APIs, the algorithm constructs a sequence of complete instructions where the adversarial command is placed in the middle, shielded by benign ``flanking'' queries. Simultaneously, it uses a text prefix to assign a fictional persona to the model. While effective, both Prompt Packer and Flanking Attack rely on heuristic template design, necessitating a shift towards mathematical optimization to automate the search for optimal injection patterns.

Addressing the inefficiency of heuristic design, Hui et al.\ (2024) introduce a deterministic optimization paradigm with \emph{PLEAK}, an automated framework for instruction-level prompt leaking~\cite{hui2024pleak}. The technical essence is ``incremental instruction optimization and encoding restoration'': utilizing gradient optimization to find the optimal adversarial query that induces the model to output system prompt fragments. As this research utilized financial datasets, its findings imply attackers can systematically steal hidden risk control logic. This marks the evolution of attacks against instruction assets from manual social engineering to automated mathematical optimization.

The synthesis reveals an evolution from static template wrapping to automated, multi-modal, and incremental optimization strategies. Financially, automated advisory platforms and multi-modal customer service bots are most exposed. These workflows are vulnerable because they reward authoritative and narrative-driven responses, providing a credible facade for masked malicious payloads.

\subsubsection{Chain/Indirect Prompt Injection}

Chain/indirect prompt injection does not touch the user/system prompt directly; instead, the payload is planted in \emph{external content} (e.g., retrieved passages or tool outputs) and, once concatenated into context, triggers boundary breaks, tool-hijacks, or data exfiltration.

Alizadeh et al.\ (2025) target tool-using agents in banking tasks, mounting chain injections that induce agents to leak personally identifiable data observed during tool execution~\cite{alizadeh2025simple}. The method constructs payloads along the data-flow of the task, implanting them in content the agent will retrieve. The key mechanism reinterprets information visible in the task from ``data'' to ``executable instructions.'' Complexity scales linearly with the number of interaction steps, $O(k \cdot m)$. However, relying on static heuristics limits the attack's resilience against active defense mechanisms such as perplexity filters.

To overcome the fragility of static injections against defensive filters, Zhan et al.\ (2025) propose an \emph{adaptive optimization framework} that systematically defeats existing indirect-injection defenses~\cite{zhan2025adaptive}. The optimization seeks higher ASR while explicitly preserving attack semantics after defense transforms (e.g., paraphrasing). The core idea is to embed defense criteria into the attack objective. Complexity grows linearly in token length and iteration steps, $O(T \cdot n)$. Nevertheless, such iterative optimization methods entail high computational overhead per sample and rely on dense feedback, rendering them inefficient for strictly black-box commercial APIs.

Addressing the constraints of black-box opacity and query budgets, Liu et al.\ (2025) shift the paradigm to an \emph{LLM-as-optimizer}\cite{yang2023large} framework with trainable memory, generating executable chain payloads without white-box access~\cite{liu2025autohijacker}. The system treats agent responses as sparse rewards and builds a trainable attack memory that transfers across samples. The key mechanism is two-stage, memory-based optimization that compensates for sparse black-box feedback. Complexity is $O(B \cdot T \cdot n)$ during training and roughly $O(n)$ at test. On commercial agents, it achieves average ASR of $\approx$71.9\%. Because it requires neither model nor tool internals, it transfers directly to financial RAG report summarization and KYC document parsing.

This subtype emphasizes planting payloads in external content to hijack tools. Entry points include RAG documents and API outputs. Financially, advisor/customer dialogues and KYC/AML toolchains are most exposed because they inherently ingest external text and possess executable interfaces, amplifying ``content-as-instruction'' risk.

\subsubsection{Comparative Analysis of Prompt Injection Subtypes}

Table~\ref{tab:prompt-injection-summary} systematically contrasts the four levels of prompt injection across algorithmic mechanisms, complexity profiles, and financial risk priority. The analysis reveals a clear escalation in systemic threat: while character (a1) and word-level (a2) injections exploit relatively shallow parsing or tokenization gaps with varying computational costs, sentence (a3) and chain (a4) injections directly compromise the high-level logic of AI agents. Chain injection represents the highest risk priority in financial contexts because it requires no direct user interaction with the prompt; instead, it leverages the model's own retrieval mechanisms (RAG) to ingest malicious payloads from external reports or web content, effectively turning the system's data ingestion pipeline into an attack vector against its own tool-use privileges.

\begin{table*}[t]
\small
\centering
\caption{Comparison of prompt injection attack levels across mechanism, computational profile, attack prerequisites, target financial workflows, and downstream deployment risk in LLM-enabled financial systems.}
\label{tab:prompt-injection-summary}
\renewcommand{\arraystretch}{1.3}
\setlength{\tabcolsep}{6pt}
\begin{tabularx}{\textwidth}{l
                            >{\raggedright\arraybackslash}X
                            >{\raggedright\arraybackslash}X
                            >{\raggedright\arraybackslash}X
                            >{\raggedright\arraybackslash}X}
\toprule
\textbf{Dimension} &
\textbf{a1 Character/Symbol} &
\textbf{a2 Word/Phrase} &
\textbf{a3 Sentence/Instruction} &
\textbf{a4 Chain/Indirect} \\
\midrule
\multicolumn{5}{c}{\textit{Algorithmic \& Methodological Contrast}} \\
\midrule
\textbf{Mechanism} &
Encoding quirks (zero-width, homoglyphs) to flip tokenization~\cite{sang2024evaluating,hackett2025bypassing}. &
Latent space semantic optimization~\cite{liu2025semantics} vs. visual homoglyphs~\cite{rizvani2026adversarial}. &
Template encapsulation~\cite{jiang2023prompt} vs. automated gradient search for system prompts~\cite{hui2024pleak}. &
Embedding payloads in external retrieval content (RAG) to hijack tool execution~\cite{alizadeh2025simple}. \\

\textbf{Complexity} &
Linear scan $O(n)$; highly efficient stackable transforms. &
High $O(N)$ for gradient methods; negligible $O(1)$ for static replacement. &
Scales with generation length $O(T \times n)$ or iterative optimisation steps. &
High training cost $O(B \cdot T \cdot n)$ for memory-based attacks; $O(n)$ at test time~\cite{liu2025autohijacker}. \\

\textbf{Prerequisites} &
Black-box access; no external tools needed. &
White-box gradients for semantic optimization; black-box for structural attacks. &
Direct prompt input; effective in black-box via transferability. &
Control over external data sources (web/docs); system prompts remain fixed. \\

\midrule
\multicolumn{5}{c}{\textit{Financial Deployment \& Systemic Impact}} \\
\midrule
\textbf{Target Workflow} &
Advisor chats, KYC forms, compliance summaries (normalization gaps). &
High-frequency trading news feeds (latency-sensitive) vs. compliance auditing. &
Automated advisory platforms and multi-modal customer service bots. &
RAG-based report summarization, KYC document parsing, and order routing agents. \\

\textbf{Risk Priority} &
Medium &
Medium &
High &
Very High \\
\bottomrule
\end{tabularx}
\end{table*}

The evolution of literature within these subcategories reflects a consistent methodological shift from static heuristics to adaptive optimization. At the character level, attacks have refined from simple visual typos to sophisticated encoding-level obfuscation designed to bypass normalization layers~\cite{hackett2025bypassing}. In the word/phrase domain, the trade-off bifurcates between computationally heavy semantic optimization for high-value targets~\cite{liu2025semantics} and lightweight structural deception for latency-critical streams~\cite{rizvani2026adversarial}. Instruction-level attacks have progressed from manual template encapsulation~\cite{jiang2023prompt} to automated gradient-based leaking~\cite{hui2024pleak}. Finally, indirect injection demonstrates the most advanced evolution, moving from static payloads in banking tasks~\cite{alizadeh2025simple} to defense-aware adaptive optimization~\cite{zhan2025adaptive} and ultimately to memory-augmented black-box automation~\cite{liu2025autohijacker}, specifically ensuring payloads survive the complex data transformations inherent in financial RAG workflows.

\subsection{Deepfake}
\label{subsec:deepfake}

While prompt injection targets the system's execution logic, deepfake attacks target the \emph{verification layer} by compromising the integrity of visual and auditory inputs. This class of attacks utilizes deep generative models, such as Generative Adversarial Networks (GANs) and Diffusion Models, to synthesize or manipulate hyper-realistic multimedia content~\cite{zhang2024diffmorpher,tolosana2020deepfakes}. The core mechanism involves modelling the manifold distribution of real biometric data to generate synthetic identities that are statistically indistinguishable from genuine samples to the target sensor. This exploits the fundamental trust assumption that financial pipelines place on audiovisual evidence as a proof of physical presence and identity.

The proliferation of deepfake technology poses a systemic threat to the foundational ``trust anchors'' of the financial ecosystem: remote identity verification and fraud detection. In Know Your Customer (KYC) pipelines, adversaries employ synthetic media to penetrate the initial trust boundary, creating synthetic ``mule'' accounts that facilitate money laundering without a physical trace. In high-value transaction authorization, high-fidelity real-time voice clones and face swaps are used to circumvent liveness detection mechanisms, granting unauthorized access to assets protected by biometric locks\cite{liu2023asvspoof}. Beyond technical bypass, deepfakes facilitate sophisticated social engineering (e.g., ``CEO fraud''), where synthetic audio-visual signals manipulate corporate decision-making hierarchies\cite{brundage2018malicious}. By decoupling digital representation from physical reality, deepfakes force financial institutions to fundamentally reconsider the validity of remote verification, escalating the cost of compliance and introducing severe friction into legitimate user interactions.

\subsubsection{Latent/Identity-Level Synthesis}
\label{subsubsec:latent-identity-synthesis}

Latent/Identity-level synthesis involves generating entirely new, non-existent facial images or manipulating the latent transition between identities by sampling from the learned manifold of generative models (e.g., StyleGAN, Diffusion), rather than modifying pixel-level noise. Unlike pixel-level perturbations, this category exploits the mathematical structure of the latent space to construct synthetic identities that possess specific demographic attributes or ``master'' patterns. In finance, this is the engine behind \emph{synthetic identity fraud} and \emph{advanced morphing}. Attackers operate in the abstract feature space to create faces that are mathematically optimized to bypass vector-based matching or duplicate detection, leaving forensic teams with synthesized ``ghosts'' rather than physical suspects.

Papantoniou et al.\ (2024) introduce \emph{Arc2Face}, shifting facial generation from text-dependency to rigorous identity control by training a projection network that maps discriminative ID embeddings (e.g., ArcFace) directly into the latent space of a frozen Stable Diffusion\cite{rombach2022high} backbone~\cite{papantoniou2024arc2face}. This decouples identity retention from style variations, forcing the generator to produce an infinite variety of photorealistic images all mapping to a single consistent identity vector, with inference scaling linearly at $O(T)$. Financially, the ability to manufacture a complete ``digital footprint'' from one synthetic ID vector directly enables scalable Synthetic Identity Fraud (SIF) bypassing cross-verification consistency checks, and because generation is driven by the same discriminative embeddings used in banking recognition, the resulting faces are inherently optimised to score high in similarity metrics.

Distinct from this deterministic identity generation, Nguyen et al.\ (2022) reveal with \emph{Master Face Attacks} that recognition systems are vulnerable to ``wolf samples'' capable of falsely matching a significant population percentage~\cite{nguyen2022master}. Using Latent Variable Evolution (LVE) driven by CMA-ES\cite{hansen2001completely}, the method iteratively searches for latent-space regions where generated face embeddings overlap with acceptance radii of multiple distinct identity clusters. Although search cost is high ($O(N_{iter} \cdot N_{pop})$), the resulting artifact is persistent. This exploitation of non-uniform embedding density critically undermines 1:N Deduplication Systems and watchlist screening in financial onboarding.

Parallel to these static point-based optimisations, Zhang et al.\ (2024) utilise \emph{DiffMorpher} for \emph{trajectory-based synthesis}, establishing smooth semantic paths between distinct identities~\cite{zhang2024diffmorpher}. The approach fits separate Low-Rank Adaptations (LoRA) to each input identity and performs dual interpolation, injecting interpolated Self-Attention keys and values during denoising to enforce low-level texture consistency at $O(T + k_{train})$ complexity. This preservation of high-frequency details transfers directly to bypassing Video-based KYC Liveness Detection, where temporal inconsistencies are the primary fraud indicator.

Latent-level attacks have evolved from optimizing error statistics (\emph{Master Faces}) to the generative engineering of consistent identities (\emph{Arc2Face}) and seamless transitions (\emph{DiffMorpher}). The technical trajectory has shifted from ``searching'' the latent space to ``projecting'' into it and ``adapting'' it via LoRA, allowing for increasingly precise manipulation of the identity vector. The primary threat landscape spans Synthetic Identity Fraud (SIF) in onboarding via consistent synthetic faces and Morphing Attacks in e-KYC via seamless diffusion interpolation. For governance, financial institutions must prioritize monitoring Embedding Density to flag vectors sitting in statistically dense clusters (potential wolves) and deploy Injection Attack Detection to verify that image streams originate from physical camera sensors rather than generative pipelines.

\subsubsection{Feature/Attribute-Level Transfer}
\label{subsubsec:feature-attribute-transfer}

Feature/Attribute-level transfer, commonly known as face swapping, relies on encoder-decoder architectures to disentangle identity features from attribute features. The attack mechanism involves forcefully injecting the identity vector of a target victim into the attribute framework of an attacker, generating a composite image that retains the attacker's context but displays the victim's biometric traits. In financial deployment, this directly targets face recognition and comparison algorithms (1:1 matching).

Zhao et al.\ (2023) introduce \emph{DiffSwap}, reformulating face swapping as controllable conditional inpainting via a ``3D-Aware Masked Diffusion'' framework~\cite{zhao2023diffswap}. Rather than relying on complex feature disentanglement, the method employs 3D face reconstruction to generate a precise semantic mask, synthesising the swapped face conditioned on the source identity vector and the target's landmarks at $O(T)$ inference complexity. By preserving the exact 3D structure and pose of the target while rendering the source identity, the technique directly neutralises Active Liveness Detection, as an attacker can perform required liveness actions in real-time video with the model faithfully rendering these movements on the victim's face.

Complementing this generative capability, Li et al.\ (2023) conduct with \emph{LiveBugger} the first systematic security assessment of commercial Facial Liveness Verification (FLV) services~\cite{li2022seeing}. The automated framework integrates face synthesis into a feedback loop: analysing the FLV API challenge prompt, synthesising corresponding victim footage performing the required action, and injecting the stream into the verification session at $O(1)$ per-frame cost. The study reveals that leading commercial FLV APIs often prioritise action correctness over source authenticity, enabling high bypass rates.

Feature-level transfer has matured from static image replacement to action-consistent video synthesis. The technical progression—exemplified by \emph{DiffSwap}'s 3D-aware shape preservation and \emph{LiveBugger}'s automated action response—demonstrates that attackers can now satisfy complex Interactive Liveness Challenges. The primary financial exposure is the Remote KYC (e-KYC) process and High-Assurance Account Recovery, where liveness is the sole barrier against identity theft. Governance strategies must shift from ``challenge-response'' to Source Integrity Verification. Priority measures include implementing Video Injection Detection (detecting virtual camera drivers), analyzing Micro-Texture Pulse Signals (rPPG) which are often lost in diffusion synthesis, and randomizing challenges with strict latency bounds to break the generation-injection loop.

\subsubsection{Adversarial/Anti-Forensic Generation}
\label{subsubsec:adversarial-anti-forensic}

Adversarial/Anti-forensic generation represents an evolution of deepfakes designed specifically to evade automated detection models deployed by financial institutions. This class of attacks integrates adversarial loss functions during the generation process or injects imperceptible gradient-based perturbations to minimize the response of known forensic detectors.

Li et al.\ (2023) extend adversarial generation to the physical sensor layer with \emph{Physical-World Optical Adversarial Attacks}, targeting structured-light 3D Face Recognition (FR) systems~\cite{li2023physical}. Instead of digitally modifying an image file, the authors propose a ``structured-light attack'' that projects carefully computed adversarial illumination patterns onto a live face. The mechanism employs a differentiable 3D reconstruction module to optimize the projected light pattern end-to-end. The optimization complexity involves simulating the physical rendering process, scaling with the resolution of the structured light pattern $O(H \cdot W)$. In the financial domain, due to its ability to manipulate depth information, this technique directly compromises Hardware-based Liveness Detection, effectively turning a spoofing attempt into a mathematically valid 3D geometry in the eyes of the sensor.

Dong et al.\ (2019) address the challenge of generating adversarial samples against black-box systems with \emph{Efficient Decision-Based Attacks}~\cite{dong2019efficient}. The authors introduce an evolutionary attack algorithm that significantly reduces the query budget required to find a successful adversarial perturbation. Mechanistically, the method models the local geometry of the search directions and employs a guided covariance matrix adaptation (CMA) to iteratively refine the noise pattern. The complexity is defined by the query count $Q$, achieving success with orders of magnitude fewer queries ($O(Q)$) than traditional gradient-estimation methods. Financially, because most third-party identity verification APIs operate as strict black boxes, this method is highly transferable to API-based Model Evasion.

Adversarial generation has bifurcated into two dangerous extremes: physical-layer manipulation (Li et al.) that targets the sensor hardware via optical projection, and logical-layer optimization (Dong et al.) that targets the decision logic via query-efficient evolution. The common thread is the move away from ``white-box'' assumptions toward realistic constraints—handling physical noise or opaque API responses. The primary financial exposure lies in Mobile Banking Authentication (vulnerable to optical 3D spoofing) and Third-Party KYC APIs (vulnerable to decision-based query attacks). Governance must evolve beyond static image analysis to include Hardware Integrity Checks (detecting abnormal illumination patterns via active sensors) and Stateful Query Monitoring (identifying evolutionary search patterns in API logs).

\subsubsection{Comparative Analysis of Deep Fake Subtypes}

\begin{table*}[t]
\small
\centering
\caption{Comparison of deepfake attack levels in financial identity and authentication systems, summarizing their underlying mechanisms, computational characteristics, threat capabilities, target financial tasks, and downstream deployment risk.}
\label{tab:deepfake-summary}
\renewcommand{\arraystretch}{1.3}
\setlength{\tabcolsep}{6pt}
\begin{tabularx}{\textwidth}{l
                            >{\raggedright\arraybackslash}X
                            >{\raggedright\arraybackslash}X
                            >{\raggedright\arraybackslash}X}
\toprule
\textbf{Dimension} &
\textbf{b1 Latent/Identity-level} &
\textbf{b2 Feature/Attribute-level} &
\textbf{b3 Adversarial-level} \\
\midrule
\multicolumn{4}{c}{\textit{Algorithmic \& Methodological Contrast}} \\
\midrule
\textbf{Mechanism} &
Deterministic latent projection~\cite{papantoniou2024arc2face} or stochastic manifold search~\cite{nguyen2022master} to create consistent new IDs. &
Identity injection via 3D-aware Diffusion~\cite{zhao2023diffswap} or automated challenge-response loops~\cite{li2022seeing}. &
Optical projection of adversarial patterns~\cite{li2023physical} or black-box query optimization~\cite{dong2019efficient}. \\

\textbf{Complexity Profile} &
Search $O(N_{iter} \cdot N_{pop})$ or Generation $O(T)$; one-time cost for persistent assets. &
Inference $O(T)$ for high-fidelity or $O(1)$ for real-time streams; requires aligned source-target geometry. &
Physical rendering $O(H \cdot W)$ or Query budget $O(Q)$; constrained by sensor resolution or API limits. \\

\textbf{Key Threat Capability} &
Creating ``Wolf'' samples (dense error centers) or consistent digital footprints for mule accounts. &
Preserving 3D structure/pose while swapping ID; bypassing ``blink/nod'' active liveness challenges. &
Creating mathematically valid 3D depth maps (sensor spoofing); Universal evasion keys for black-box APIs. \\

\midrule
\multicolumn{4}{c}{\textit{Financial Deployment \& Systemic Impact}} \\
\midrule
\textbf{Target Financial Task} &
Automated Onboarding (1:N deduplication) and Document verification (SIF). &
Interactive Authentication (1:1 matching) and Account Recovery (ATO). &
Hardware Sensors (Depth/IR) for mobile login and Black-box Fraud APIs. \\

\textbf{Risk Priority} &
High &
Very High &
Medium--High \\
\bottomrule
\end{tabularx}
\end{table*}

Table~\ref{tab:deepfake-summary} systematically compares the three deepfake subtypes across algorithmic mechanisms, complexity profiles, and financial risk priority. The unifying objective across all categories is to decouple the semantic identity from the physical source, thereby invalidating the ``what you see is who is there'' assumption of remote verification. Feature-level transfer (Face Swapping) presents the highest systemic risk priority because it directly neutralizes active liveness detection—the primary firewall for high-value transactions—by preserving 3D geometry and micro-expressions under interactive challenges, enabling scalable Account Takeover (ATO)~\cite{zhao2023diffswap,li2022seeing}. Latent-level synthesis (Identity Generation) facilitates Synthetic Identity Fraud (SIF) by manufacturing consistent, non-existent identities (``ghosts'') or exploiting error-dense regions (``wolves''), effectively polluting the KYC database with mule accounts that evade deduplication~\cite{papantoniou2024arc2face,nguyen2022master}. Adversarial generation targets the sensor or logic layer directly, using optical patterns or query optimization to bypass hardware checks or black-box APIs, though its scalability is often constrained by physical access or query limits compared to the purely digital nature of the other two~\cite{li2023physical,dong2019efficient}.

The evolution of literature within each subclass illustrates a clear trajectory from static generation to dynamic evasion. At the latent level, the methodology shifts from stochastic searches for universal error patterns (Master Faces)~\cite{nguyen2022master} to the deterministic projection of specific identities (Arc2Face)~\cite{papantoniou2024arc2face} and finally to seamless semantic interpolation for dynamic morphing (DiffMorpher)~\cite{zhang2024diffmorpher}. In the feature-level domain, the focus advances from high-fidelity static synthesis (DiffSwap)~\cite{zhao2023diffswap} to systematic evasion frameworks capable of automating responses to interactive liveness challenges (LiveBugger)~\cite{li2022seeing}. Parallel to these, adversarial generation evolves from physical-layer manipulation of sensor hardware via optical projection~\cite{li2023physical} to logical-layer optimization targeting black-box decision boundaries via query-efficient algorithms~\cite{dong2019efficient}, collectively demonstrating that the threat surface has expanded from simple visual spoofing to the mathematical compromise of the entire authentication stack.

\section{Literature Review Comparison and Positioning}
\label{sec:lit-review}
Recent survey papers have examined adjacent aspects of AI, cybersecurity, and trustworthiness in finance, but they differ substantially in scope, granularity, and target problem definition. To clarify the positioning of this survey, we compare our work against six representative surveys spanning (i) FinTech/digital banking cybersecurity, (ii) broad AI-in-finance adoption and governance, (iii) financial explainable AI (XAI), and (iv) general adversarial machine learning (AML) reviews/meta-surveys.

\subsection{(1) FinTech and digital banking cybersecurity surveys}

Javaheri et al. (2024) provide a systematic review of cybersecurity threats in FinTech, with a strong emphasis on threat taxonomy, attack categories, and defense strategies across the FinTech ecosystem \cite{javaheri2024cybersecurity}. Their survey is valuable for understanding the broader cyber threat landscape in financial technology systems. However, its primary focus is cybersecurity at the system/platform level (e.g., threats, vulnerabilities, and controls in FinTech infrastructures), rather than the security and robustness properties of AI models deployed in financial decision-making pipelines.

Similarly, Asmar and Tuqan (2024) review the use of machine learning for sustaining cybersecurity in digital banks, highlighting how ML techniques can be applied to detect or mitigate banking cyber threats (e.g., anomaly detection, fraud-related patterns, or network/security incidents)~\cite{asmar2024integrating}. This line of work is highly relevant to banking operations, but it mainly treats ML as a tool for cybersecurity. In contrast, our survey focuses on the inverse perspective: the security and robustness of finance AI systems themselves, including how model-level vulnerabilities (e.g., poisoning, evasion, distribution shift, model extraction/privacy leakage, unreliable inference behavior) affect financial applications and risk control.

\subsection{(2) Broad AI-in-finance surveys with governance/regulatory focus}

Vuković et al. (2025) survey AI integration in financial services from a broader perspective, covering adoption trends, use cases, and regulatory/ethical challenges. Such surveys are important for understanding the macro-level trajectory of AI in finance, including governance, compliance, and institutional barriers to deployment~\citep{vukovic2025ai}. However, their treatment of security and robustness is typically one component among many (alongside adoption, regulation, explainability, and ethics), and does not provide a dedicated framework for systematically analyzing AI-specific attack surfaces, failure modes, and robustness evaluation protocols across the financial AI lifecycle.

Our survey complements this literature by narrowing the focus to security and robustness as first-class design and deployment concerns in finance AI. Rather than primarily asking where AI is used in finance or how regulation shapes adoption, we ask: how and where finance AI systems fail under adversarial, uncertain, or non-stationary conditions; how these failures propagate into financial decisions; and how such risks should be evaluated and governed.

\subsection{(3) Financial XAI and trustworthy AI surveys}

Yeo et al. (2025) provide a comprehensive review of explainable AI in finance. This is a crucial neighboring literature because explainability is often treated as a pillar of trustworthy finance AI~\cite{yeo2025comprehensive}. Their survey addresses methods, use cases, and challenges in making financial AI decisions interpretable and auditable.

Nevertheless, explainability and robustness address different (though related) questions. XAI-focused surveys primarily ask whether a model’s outputs can be interpreted, justified, or audited, whereas our survey asks whether the model remains secure, stable, and reliable under adversarial manipulation, data quality degradation, distribution drift, and deployment-time uncertainty. In practice, a model may be explainable yet fragile, or robust in some conditions yet opaque. We therefore position this survey as complementary to financial XAI surveys by emphasizing the security/robustness axis of trustworthy finance AI.

\subsection{(4) General adversarial machine learning surveys and meta-surveys}

Pelekis et al. (2025) review adversarial machine learning methods, tools, and critical industry sectors \cite{pelekis2025adversarial}, while Pawlicki et al. (2025) provide a meta-survey of adversarial attacks across AI domains (including emerging model classes) \cite{pawlicki2025meta}. These surveys are highly relevant from a methodological perspective because they synthesize attack/defense taxonomies, benchmarks, and evaluation practices across domains.

However, general AML surveys are necessarily domain-agnostic. They usually do not fully capture the finance-specific operational constraints and consequences that shape practical robustness requirements, such as:

\begin{itemize}
    \item real-time execution constraints in trading and fraud detection,
    \item asymmetric error costs (e.g., false negatives vs.\ false positives under regulatory obligations),
    \item auditability and compliance requirements,
    \item market/portfolio propagation effects,
    \item systemic risk amplification due to correlated model failures,
    \item and institution-specific governance/control processes.
\end{itemize}

Our survey builds on the AML literature but re-organizes the problem for the financial context. Specifically, we translate general AML and robustness issues into a finance-centered, lifecycle-aware framework that links technical vulnerabilities to application-level and system-level financial risks.

\subsection{How this survey differs from prior surveys}

Relative to the above literature, this survey is distinguished by four design choices.

First, finance AI security and robustness is the central object of study, rather than a subtopic within broader AI-in-finance, cybersecurity, or XAI surveys. This allows us to provide deeper coverage of failure mechanisms and defenses specific to AI-driven financial systems.

Second, we adopt an organizational framework that combines lifecycle stages with mechanism-level distinctions, tracing risks across the full financial AI pipeline from data and model development to deployment, monitoring, and governance, while further differentiating distinct vulnerability mechanisms and attack forms within each stage. This helps unify threats and mitigations that are often studied in isolation. This is particularly important in finance, where risk may be introduced upstream (e.g., data quality or labeling issues) but only materialize downstream (e.g., trading losses, compliance breaches, or biased credit decisions).

Third, we explicitly connect mechanism-level vulnerabilities to financial consequences. Beyond cataloging attacks and defenses, we emphasize how technical failures propagate into business, regulatory, and systemic outcomes in financial settings. This connection is often underdeveloped in both general AML reviews and broad financial AI surveys.

Fourth, we highlight evaluation and governance gaps for high-stakes financial deployment, including the need for finance-relevant robustness benchmarks, stress testing protocols, post-deployment monitoring standards, and governance integration between ML teams, risk management, compliance, and audit functions.

Taken together, our survey is intended to bridge a gap between (i) domain-agnostic adversarial/robustness surveys and (ii) broad financial AI application/governance surveys, by providing a dedicated, finance-specific synthesis of AI security and robustness challenges, methods, and open research directions.

\section{Conclusion}
\label{sec:conclusion}

Financial AI systems increasingly operate as continuously deployed, automation-amplified pipelines in which small algorithmic perturbations can propagate through long decision chains and persist via feedback loops. In this survey, we synthesized the security and robustness literature in finance AI through a lifecycle-centric and mechanism-driven lens, motivated by the mismatch between domain-agnostic adversarial machine learning taxonomies and the operational realities of financial deployment.

We organized the threat landscape across three lifecycle stages: training and updating, where poisoning of data and model updates can embed persistent vulnerabilities into learned decision logic; deployment and inference, where evasion and transfer-based attacks exploit decision-boundary sensitivity under realistic interaction constraints; and operation, monitoring, and feedback, where workflow-level manipulation, including prompt injection against LLM-based financial assistants and deepfake-driven compromise of KYC/eKYC layers, can induce delayed, cascading, and hard-to-audit harm. Across these stages, we emphasized mechanism-level commonalities such as recursive amplification, statistical or gradient alignment, transferability, and distributional drift, while contextualizing feasibility and downstream impact under finance-specific constraints including real-time execution, asymmetric error costs, compliance obligations, and automation-amplified downstream effects.

Beyond categorizing attacks, this survey identifies several open challenges and under-explored directions for securing financial AI pipelines. First, persistent manipulation across updates remains insufficiently understood, especially in settings involving continuous retraining, federated aggregation, and long-horizon feedback loops, where small adversarial effects may accumulate into delayed systemic degradation. Second, the robustness and integrity of monitoring signals themselves require greater attention, since post-deployment safeguards may be bypassed, delayed, or corrupted by adaptive adversaries, undermining anomaly detection, auditability, and human oversight. Third, emerging operational threats in AI-mediated workflows, including LLM-agent prompt injection, retrieval-chain hijacking, and deepfake-enabled identity subversion, expand the attack surface beyond model parameters and decision boundaries into the workflow logic and evidentiary interfaces that increasingly mediate financial decisions.

These gaps point to a broader research agenda. Securing financial AI pipelines will require moving beyond isolated attack demonstrations toward lifecycle-aware stress testing, finance-relevant robustness benchmarks, robust monitoring and response loops, and governance integration across ML teams, risk management, compliance, and audit functions. We hope this survey provides a unifying foundation for researchers and practitioners to understand, compare, and mitigate adversarial threats in finance AI systems, and to develop security and robustness methods that are aligned with real financial deployment constraints and high-stakes operational requirements.

\bibliographystyle{ACM-Reference-Format} %
\bibliography{fintechsurvey} %

\end{document}